\documentclass[runningheads]{svmult}
\usepackage{graphicx}  
\usepackage{subeqnar}  
\usepackage{multicol}  
\makeindex             
\usepackage{feynmp,epsfig,amsfonts,amsbsy}

\newcommand{\ii}{{\mathrm i}}

\renewcommand{\Im}{{\mathrm{Im}}}
\newcommand{\lsim}{\stackrel{\scriptstyle <}{\phantom{}_{\sim}}}
\newcommand{\gsim}{\stackrel{\scriptstyle >}{\phantom{}_{\sim}}}
\def\lsim{\unitlength5mm\begin{picture}(1,0)
\put(0,.35){\makebox(1,0){$<$}}\put(0,0){\makebox(1,0){$\sim$}}
\end{picture}}
\def\eqs{\makebox(0,0){$=$}}
\def\pls{\makebox(0,0){$+$}}

\def\ssp{\makebox(0,0)
    {\thinlines\put(-.1,0){\line(1,0){.2}}\put(0,-.1){\line(0,0){.2}}}}
\def\ssm{\makebox(0,0){\put(-.1,0){\thinlines\line(1,0){.2}}}}
\def\photon{\thinlines\multiput(0,0)(.2,0){3}{\line(1,0){0.1}}}
\def\lphoton{\thinlines\multiput(0,0)(.2,0){5}{\line(1,0){0.1}}
\put(0.4,0){\vector(1,0){0.1}}}

\def\oneloop{
     \put(1.5,0){\thicklines\oval(2.0,1.5)}
     \put(0,0){\photon}\put(0.3,0.3){\ssp} 
     \put(2.5,0){\photon}\put(2.7,0.3){\ssm}}
\def\fullself{\begin{picture}(5,1)\put(0,0){\oneloop}
     \put(1.5,0){\makebox(0,0){$-\ii{\boldsymbol \Pi}$} }
     \end{picture}}

\def\oneloopvertex{
    \put(1.625,0){\thicklines\oval(2.0,1.5)}
    \put(0,0){\photon}\put(0.35,0.3){\ssp} 
    \put(0.625,0){\circle*{.25}}\put(2.625,0){\circle*{.25}}
    \put(2.75,0){\photon}\put(2.9,0.3){\ssm}}
\def\Oneloopvertex{
    \put(1.625,0){\thicklines\oval(2.0,1.)}
    \put(0,0){\photon}\put(0.35,0.3){\ssp} 
    \put(0.625,0){\circle*{.25}}\put(2.625,0){\circle*{.25}}
    \put(2.75,0){\photon}\put(2.9,0.3){\ssm}}
\def\mediumloop{
    \put(2.125,0){\thicklines\oval(3.0,1.5)}
    \put(0,0){\photon}\put(0.35,0.3){\ssp} 
    \put(0.625,0){\circle*{.25}}\put(3.625,0){\circle*{.25}}
    \put(3.75,0){\photon}\put(3.9,0.3){\ssm}}

\def\doubleloop{
    \put(1.125,0){\thicklines\oval(1,2)}\put(2.625,0){\thicklines\oval(1,2)}
    \put(0,0){\photon}\put(0.35,0.3){\ssp} 
    \put(0.625,0){\circle*{.25}}\put(3.125,0){\circle*{.25}}
    \put(3.25,0){\photon}\put(3.4,0.3){\ssm}
    \multiput(1.875,-.4)(0,.8){2}{\makebox(0,0){\rule{3mm}{1.5mm}}}
    \put(1.875,-.8){\ssm}\put(1.875,.8){\ssp} }
\def\Doubleloop{
    \put(1.125,0){\thicklines\oval(1,2)}\put(3.625,0){\thicklines\oval(1,2)}
    \put(0,0){\photon}\put(0.35,0.3){\ssp} 
    \put(0.625,0){\circle*{.25}}\put(4.125,0){\circle*{.25}}
    \put(4.25,0){\photon}\put(4.4,0.3){\ssm}
    \multiput(1.875,-.4)(0,.8){2}{\makebox(0,0){\rule{3mm}{1.5mm}}}
    \multiput(2.875,-.4)(0,.8){2}{\makebox(0,0){\rule{3mm}{1.5mm}}}
    \multiput(2.375,-.4)(0,.8){2}{\thicklines\oval(.5,.7)}
    \put(1.875,-.8){\ssp}\put(1.875,.8){\ssp} 
    \put(2.875,-.8){\ssm}\put(2.875,.8){\ssm} }

\def\borndiag{\begin{picture}(2.5,1.5)\thicklines     
     \put(.25,.25){\vector(1,0){.75}}
     \put(1.125,.5){\makebox(0,0){\rule{2mm}{4mm}}}
     \put(1.,.25){\vector(1,0){1.75}}
     \put(.25,.75){\vector(1,0){.75}}\put(1.,.75){\vector(1,0){.9}}
     \put(2.,.75){\circle*{.2}}\put(2.,.75){\vector(1,0){.75}}
     \multiput(2.,.75)(0,.25){3}{\thinlines\line(0,1){.15}}\end{picture}}
\def\twocol{\begin{picture}(2.5,1.5)\thicklines     
     \put(.25,0){\vector(1,0){.75}}
     \put(1.125,.375){\makebox(0,0){\rule{2mm}{6mm}}}
     \put(1.,0){\vector(1,0){.8}}
     \put(.25,.75){\vector(1,0){.75}}\put(1.,.75){\line(1,0){1}}
     \put(1.35,0){\borndiag}\end{picture}}
\def\selfinsert{\begin{picture}(5,2)\put(0,0){\Oneloopvertex}
     \put(1.2,.688){\fullbox}
     \put(1.2,.35){\ssp}\put(2.05,.35){\ssm}\put(2.05,.688){\fullbox}
     \put(.95,.875){\ssp}\put(2.3,.875){\ssm}
     \thicklines\put(1.625,.875){\oval(1.,.6)[t]}
     \put(1.2,.875){\line(1,0){.9}}
     \put(1.575,.5){\vector(1,0){.2}}
     \put(1.575,1.175){\vector(1,0){.2}}
     \put(1.675,-.5){\vector(-1,0){.2}}\put(1.675,.875){\vector(-1,0){.2}}
     \end{picture}}

\def\fullbox{\makebox(0,0){\rule{1.5mm}{3mm}}}
\def\interaction{\makebox(0,0){\put(0,0){\interact}
    \put(0,.95){\ssp}\put(0,-.95){\ssm}
    \put(0,.125){\ssp}\put(0,-.125){\ssm}}}
\def\interact{\makebox(0,0){\put(0,.5){\fullbox}
    \thicklines\put(0,0){\oval(.75,.5)}
    \put(0,-.5){\fullbox}} }
\def\intrad{\begin{picture}(3,1.5)\put(1,.75){\interact}\thicklines
     \put(0,0){\vector(1,0){.9}}\put(0,1.5){\vector(1,0){.9}} 
     \put(.9,0){\vector(1,0){1.1}}\put(.9,1.5){\vector(1,0){1.1}}
 \put(1.375,.75){\circle*{.3}}\put(1.375,.75){\lphoton}\end{picture} }
\def\til2loop{\put(0,0){\oneloopvertex}\put(4.,0){\pls}
      \put(4.75,0){\oneloopvertex}\put(6.375,0){\interaction}
      \put(9,0){\pls}}

\def\keydiagrams{\begin{picture}(21,4)\put(0,3){
   \put(0,0){\fullself}\put(3.5,0){\eqs}\put(4,0){\til2loop}}
   \put(13.75,3){\put(0,0){\mediumloop}
      \multiput(1.625,0)(1,0){2}{\interaction}}
      \put(18.5,3){\pls}\put(19.1,3){\makebox(0,0){$\dots$}}
   \put(-3,0){
   \put(4,.2){\doubleloop}\put(8.5,.2){\pls}\put(3.25,.2){\pls}
   \put(9.25,.2){\put(0,0){\mediumloop}
      \thicklines\multiput(1.615,-.75)(.02,0){10}{\line(2,3){1}}
      \multiput(2.615,-.75)(.02,0){10}{\line(-2,3){1}}
      \put(1.625,.95){\ssp}\put(1.625,-.95){\ssm}
      \put(2.625,.95){\ssm}\put(2.625,-.95){\ssp}}
   \put(14.25,.2){\pls}\put(15,.2){\Doubleloop}
   \put(20.5,.2){\pls}\put(21.5,.2){\makebox(0,0){$\dots$}}}
   \end{picture}}

\makeatletter
\begin{fmffile}{medneutrd}
\begin{document}
\title*{Neutrino Cooling of Neutron 
Stars. Medium effects\protect
\footnote{To be published in "Physics of Neutron Star Interiors", 
Eds. D. Blaschke, N.K. Glendenning,
A. Sedrakian, Springer, Heidelberg (2001)}
}
\titlerunning{Medium Effects }
\author{Dmitri~N.~Voskresensky}
\authorrunning{D.N.~Voskresensky}
\institute{Moscow Institute for Physics and
Engineering, Russia, 115409 Moscow, Kashirskoe shosse 31;
        Gesellschaft f\"ur Schwerionenforschung GSI,
        P.O.Box 110552, D-64220 Darmstadt, Germany}
\maketitle
\begin{abstract} 
This review demonstrates that
neutrino emission from dense hadronic component in neutron stars
is subject of strong modifications due to collective effects in the
nuclear matter. With the most important in-medium processes
incorporated in the cooling code an overall agreement with 
available soft $X$ ray data can be 
easily achieved. With these findings so called
{\em{"standard"}} and {\em{"non-standard"}} cooling scenarios are replaced
by one general {\em{"nuclear medium cooling 
scenario"}} which relates slow and rapid neutron
star coolings to the star masses (interior densities). 
In-medium effects take important part also
at early hot stage of  neutron star  evolution decreasing the neutrino
opacity for less massive and increasing for more massive neutron stars.  
A formalism for calculation of neutrino radiation
from nuclear matter is presented that treats on equal footing 
one-nucleon and multiple-nucleon processes as well as reactions
with resonance bosons and condensates. 
Cooling history of neutron stars with quark cores is also discussed.
\end{abstract}


\section{Introduction}
The {\small EINSTEIN}, {\small EXOSAT} and {\small ROSAT}
observatories have measured surface
temperatures of certain neutron stars (NS) and put upper limits on the
surface temperatures of some other NS
( cf. \cite{ST83,PTT92,AKP95} and further references therein).
The data for some supernova
remnants indicate rather
slow cooling, while the data for several pulsars 
point to an essentially more rapid cooling.

Physics of NS cooling is based on a number of
ingredients, among which the neutrino emissivity of the high density
hadronic matter in the star core plays a crucial role.
Neutron star temperatures
are such that, except first minutes--hours, neutrinos/antineutrinos
radiate energy directly from the star without subsequent
collisions, since $\lambda_{\nu},
\lambda_{\bar{\nu}}\gg R$, where $\lambda_{\nu},
\lambda_{\bar{\nu}}$ are the neutrino and antineutrino mean free
paths and $R$ is star radius.  
In the so called {\em{"standard 
scenario"}} of the NS cooling (scenario
for slow cooling)
the most important channel up to temperatures $T\sim 10^{9}$K
belongs to the modified Urca (MU) process
$n \, n \rightarrow n \, p\, e\, \bar \nu$.  First estimates of its
emissivity were done in \cite{BW65,TC65}.
References \cite{FM79,M79} recalculated the emissivity of this
process in the model, where the nucleon-nucleon (NN) interaction
was treated with the help of slightly modified free one-pion exchange (FOPE).
This important result for the emissivity,
$\varepsilon _{\nu}[\mbox{FOPE}]$, was proved to be by an order of magnitude
larger than the previously obtained one. 
Namely the value
$\varepsilon _{\nu}[\mbox{FOPE}]$ was used in various
computer simulations resulting in the 
{\em{"standard scenario"}} of the cooling,
e.g. cf. \cite{T79,NT81,SWWG96}. 
Besides the MU process, in the framework of the {\em{"standard
scenario"}} numerical codes included
also processes of the nucleon (neutron and proton) bremsstrahlung (NB)
$n\, n\rightarrow n\, n \nu  \bar{\nu}$ and
$n\, p\rightarrow n\, p \nu  \bar{\nu}$, which lead
to a smaller
contribution to the emissivity then the MU, cf. 
\cite{FSB75,FM79}. Medium effects enter the above two-nucleon 
(MU and NB) rates mainly
through the effective mass of the nucleons which has a smooth 
density dependence. Therefore within FOPE model
the density
dependence of the reaction rates is rather weak and the neutrino radiation from
a NS depends 
only very weakly on its mass.  This is the
reason why the {\em{"standard scenario"}} based on the result 
\cite{FM79}, though
complying well with several slowly cooling pulsars, fails to explain the
data of the more rapidly cooling ones.
Also  {\em{"standard scenario"}} included
processes contributing to the emissivity in the NS crust
which become important at a lower temperature.

The {\em{non-standard scenario}}
included so called exotica, associated with different types of direct 
Urca-like processes, i.e.
the pion Urca (PU) \cite{MBCDM77} and kaon Urca (KU) \cite{BKPP88,T88} 
$\beta$--decay processes and direct Urca (DU) on
nucleons and hyperons \cite{LPPH91} possible 
only in sufficiently dense interiors of rather massive
NS.  The main difference in the
cooling efficiency driven by the DU-like processes on one hand
and the MU and NB processes on the other hand
lies in the rather
different phase spaces associated with these reactions.  
In the MU and NB
case the available phase space is that of a two-fermion origin, 
while in the pion (kaon) $\beta$--decay
and DU on
nucleons and hyperons it is that of a one-fermion origin.
Critical density of pion condensation in NS matter is 
$\varrho_{c\pi} 
\simeq (1\div3)\varrho_0$ depending on the type of condensation 
(neutral or charged)
and the model, see \cite{MSTV90,TT97,APR98,SST99}. 
Critical density of kaon condensation is $\varrho_{cK} 
\simeq (2\div6)\varrho_0$ depending on the type ($K^-$ or $\bar{K}^0$, 
$S$ or $P$ wave) and the model, see  \cite{BLRT94,KVK95}. 
Critical density for the DU
process is $\varrho_{cU} 
\simeq (2\div6)\varrho_0$ depending on the model for the equation of state
(EQS), 
see \cite{LPPH91,APR98}.
Recent calculations \cite{SST99}
estimated critical density of neutral pion condensation as $2.5\varrho_0$ 
and  for the charged one as $1.7\varrho_0$, whereas
variational calculations
\cite{APR98} argued even for smaller critical densities ($\simeq 1.3 \varrho_0$
for $\pi^0$ condensation). On the other hand the EQS of 
\cite{APR98}
allows for
DU process only at $\varrho >5\varrho_0$. 

There is no bridge between {\em{"standard"}} and {\em{"non-standard"}}
scenarios due to complete ignorance of in-medium modifications of $NN$ 
interaction which allows for
strong polarization of the 
soft modes (like virtual dressed
pion and kaon modes serving a part of in-medium baryon--baryon
interaction). {\em{Only due to enhancement of a medium
polarization with the baryon density pion
condensate may appear and it seems thereby quite inconsistent to ignore the 
softening effects for 
$\varrho<\varrho_{c\pi}$, 
and suddenly switch on the condensate for 
$\varrho >\varrho_{c\pi}$.}}

Now let us basing on the results \cite{VS84,VS86,VS87,SV87,MSTV90,SVSWW97}
briefly discuss a general {\em{"nuclear medium cooling 
scenario"}} 
which treats obvious caveats of  two mentioned above scenarios.
First of all one observes \cite{VS84} that in the nuclear matter 
many new reaction channels are opened up compared to
the vacuum processes. 
Standard Feynman
technique fails to calculate in-medium reaction rates if the particle widths
are important since there are no
free particle asymptotic states in matter. Then
summation of all perturbative Feynman 
diagrams where free Green functions are replaced by the  
in-medium ones leads to
a double counting due to multiple repetitions of some processes
(for an extensive discussion of this defect see
\cite{KV99}). This calls
a formalism dealing with closed diagrams (integrated over all
possible in-medium particle states) with full non-equilibrium Green 
functions. 
Such a formalism was elaborated in \cite{VS86,VS87}  first
within quasiparticle approximation (QPA)
for nucleons and was called in \cite{VS87}
{\em{"optic theorem formalism ($\rm OTF$)
in non-equilibrium
diagram technique"}}. It was demonstrated that standard
calculation of the rates via squared reaction matrix elements and calculation
using OTF coincide within QPA picture 
for the fermions. 
In  \cite{KV95}
the formalism  was generalized to include 
arbitrary particle widths effects.  
The latter formalism treats on equal footing 
one-nucleon and multiple-nucleon processes as well as resonance reaction
contributions of the boson origin, as processes with
participation of zero sounds and reactions on the boson condensates.
Each diagram in the  series with full Green functions is free from the
infrared divergences.  Both, the correct quasiparticle (QP) and
quasiclassical limits are recovered.

Except for very early stage of NS evolution (minutes - hours)
typical averaged lepton energy ($\gsim T$)
is larger then the nucleon particle width
$\Gamma_N \sim T^2/\varepsilon_{FN}$ and the nucleons can be treated within the
QPA. This observation
much simplifies consideration since
one can use an intuitive way of separation of the processes according
to the available phase space. The one-nucleon
processes have the largest emissivity 
(if they are not forbidden by energy-momentum conservations),
then two-nucleon processes come into play, etc.

\begin{figure}
\centering\psfig{figure=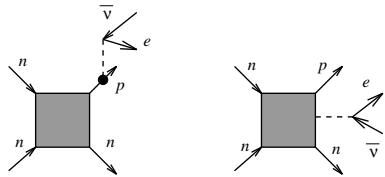,width=5cm}
\caption[]{Antineutrino emission from a nucleon leg (left graph) and 
  from intermediate scattering states (right) in MMU process. 
Full dot includes weak coupling
vertex renormalization.
 \label{fig:graph1}}
\end{figure}

In the temperature interval $T_{c}<T<T_{opac}$ ($T_c$ is typical temperature
for the nucleon pairing and $T_{opac}$ is typical temperature  
at which
neutrino/antineutrino mean free path $\lambda_\nu$/$\lambda_{\bar{\nu}}$
is approximately equal to the star radius $R$) 
the neutrino emission 
is dominated by the medium modified Urca (MMU) and
medium nucleon bremsstrahlung (MNB) processes if one-nucleon reactions like
DU, PU and KU are
forbidden, as it is the case for 
$\varrho <\varrho_{cU}, \varrho_{c\pi},\varrho_{cK}$. Corresponding diagrams 
for MMU process are
schematically shown in Fig. \ref{fig:graph1}.
References \cite{VS84,VS86,VS87,SV87,MSTV90} 
considered $NN$ interaction within Fermi liquid Landau--Migdal approach.
They incorporated the softening of the medium one-pion
exchange (MOPE) mode, other medium polarization effects, like 
nucleon-nucleon correlations in the vertices, renormalization of the
local part of $NN$ interaction by the loops, as well as
the possibility of the neutrino emission from the intermediate reaction
states and resonance DU-like reactions going on zero sounds and the 
boson condensates. 
References \cite{VS86,SV87,MSTV90} have demonstrated that for 
$\varrho \gsim \varrho_0$
second diagram of Fig. \ref{fig:graph1} gives the main contribution to the
emissivity of MMU process rather than the first one which contribution has been
earlier evaluated 
in the framework of FOPE model in \cite{FM79}.
This fact essentially modifies the absolute value as well as the
density dependence of the $nn\rightarrow npe\bar{\nu}$ 
process rate which becomes 
to be very sharp. Thereby,
for stars of masses larger than the solar mass the resulting
emissivities were proved to be substantially larger
than those values calculated in FOPE model. With increase of
the star mass (central density) pion mode continues to soften and MMU
and MNB rates still increase. At $\varrho >\varrho_{c\pi}$ 
pion condensation begins 
to contribute. Actually, the condensate droplets may be exist already at 
a smaller density  if the mixed phase is
appeared, as suggested in \cite{G92}. At $T>T_{melt}$,
where $T_{melt}$ is the melting temperature, roughly $\sim$ several MeV, 
the mixed phase is in liquid state and PU processes on independent condensate
droplets
are possible. At $T<T_{melt}$
condensate droplets are placed in a crystalline
lattice that substantially suppresses corresponding neutrino processes.

Reference \cite{MBCDM77} considered the reaction channel
$n\rightarrow p\pi^{-}_{c}e\bar{\nu}$,
whereas \cite{VS84,VS86} included other possible
pion $\pi^+$, $\pi^{\pm}$, $\pi^0$ condensate processes with charged and
neutral currents (e.g.,
like $n\pi^{0}_c \rightarrow pe\bar{\nu}$,
$n\pi^{+}_c \rightarrow p\nu\bar{\nu}$
and $n\pi^{0}_c  
\rightarrow n\nu\bar{\nu}$) as well as resonance reactions
going on zero sounds which are also possible at $\varrho <\varrho_{c\pi}$.
Due to 
$NN$ correlations all 
pion condensate rates are significantly suppressed 
(by factors $\sim$ 10 -- 100 compared to first estimate \cite{MBCDM77},
see \cite{T83,VS84,VS86,UNTMT94}. 
At $\varrho\sim \varrho_{c\pi}$ both MMU and PU
processes are of the same order of magnitude \cite{VS86}
demonstrating a smooth transition
to higher densities (star masses) 
being absent in the {\em{"standard"}} and {\em{"non-standard"}} scenarios.

\begin{figure}
\centering\psfig{figure=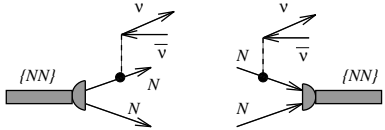,width=5cm}
\caption[]{Neutrino--antineutrino emission from Cooper pair-breaking (left 
graph)
  and pair-formation processes (right graph). 
 \label{fig:graph2}}
\end{figure}

For $T<T_c$ the reactions 
of neutrino pair radiation from superfluid nucleon pair 
breaking and formation 
(NPBF) shown in 
Fig. \ref{fig:graph2} become to be dominant processes.
The neutron pair  breaking  and formation (nPBF) 
process for the case of the $1S_0$ pairing
was first calculated in 
\cite{FRS76} using standard Bogolyubov technique. Later this process
was independently calculated in
\cite{VS87,SV87} as demonstration of efficiency of
OTF within the closed non-equilibrium
diagram technique developed 
there. Moreover  \cite{VS87} calculated emissivity of the 
corresponding process on proton (pPFB) taking into
account strong coupling $p\nu\bar{\nu}$ vertex renormalization
(see first diagram (\ref{p-vert}) below).
It results in one-two order of magnitude enhancement  
of pPFB emissivity (for $\varrho \sim (1.5\div 3)\rho_0$)
compared to that would be
estimated with the vacuum vertex,
leading to that both nPFB and pPFB
emissivities can be of the same 
order of magnitude.  
Emissivities of NPBF processes  have the same suppression 
factor $\sim \exp (-2\Delta/T)$ as MU, NB, MMU and MNB at $T<T_c$
but compared to the latter the NPBF processes have a large
one nucleon phase space
volume. Ref. \cite{VS87} and then \cite{MSTV90}
also sketched how to incorporate
$3P_2$ pairing (the developed formalism easily allows to do it) 
indicating that  exponential suppression of the
specific heat and
the emissivity is to be replaced by only a power law suppression
when the gap
vanishes at the Fermi sphere poles. This is precisely the case for the
$3P_2 (\mid m_J \mid =2)$ pairing, where $m_J$ is projection of the total pair
momentum onto quantization axis. This idea was then
worked out  in \cite{LY94,YKL99,YLS99}, without
reference.

I would also like to do a historic remark relating to 
estimation of the NPBF
processes since in some  works (e.g., see \cite{P98,YLS99}) 
was expressed a surprise  
why these processes were not 
on a market during many years. 
First work \cite{FRS76}, although found correct analytic 
expression for 
$1S_0$ pairing of neutrons, 
numerically underestimated the emissivity by an order of
magnitude. 
Also there was no  statement on the  dominance of the process
in the cooling scenario (i.e. over the MU). 
Asymptotic behaviour of the emissivity
$\varepsilon [\mbox{nPBF}]\sim 10^{20} T^7_{9}\mbox{exp}({-2\Delta/T})$
for $T\ll \Delta$, $T_9 =T/10^9 \mbox{K}$,
as follows from expression (1b) of \cite{FRS76} and from their
rough asymptotic estimate of the
integral (see below (13b)), shows up nor a large one nucleon phase space factor
($\sim 10^{28}$) nor appropriate temperature behaviour.  
Namely this underestimation of the rate in \cite{FRS76}
(we again point out that 
analytic expression (1a) is correct) and absence of
mentioning of possible dominance of the process over MU,
became the reason that this important result was overlooked during many years.
Reference \cite{VS87} overlooked sign of anomalous diagram for $S$
pairing and 
included $\propto g_A^2$ term ($g_A$
is axial-vector coupling constant) 
which should  
absent for $S$
pairing giving, nevertheless, the main contribution 
for the $3P_2$ pairing case. A reasonable 
numerical estimate  of the emissivity was presented valid
for both $S$ and $P$ pairings (except $\mid m_J \mid=2$ case) including 
$NN$ correlation effects into consideration. Uncertainty of this
estimate is given  by a factor ($0.5\div2$) which is 
allowed by variation of 
not too well known correlation factors. Namely 
this estimate was then 
incorporated within the cooling code in \cite{SVSWW97}.
Correct asymptotic behaviour of the emissivity is 
$\varepsilon [\mbox{nPFB}, \mbox{pPFB}]\sim 10^{28} 
(\Delta /\mbox{MeV})^{7}(T/\Delta )^{1/2} \mbox{exp}({-2\Delta/T})$
for $T\ll \Delta$, that showed up a huge one-nucleon phase space factor and
very moderate $T$ dependence of the pre-factor. 
Thereby the value of the rate was related in \cite{VS87} to the value of
the pairing gap.
The possibility of the  dominant role of this process 
(even compared to enhanced
MMU and PU rather than only to MU) was unambiguously stressed.
Unfortunately  \cite{VS87}
had a number of obvious misprints which  were partially corrected in
subsequent papers. Although it does not excuse
the authors these misprints   
can be easily treated by an
{\em{attentive}} reader.  Importance of NPFB processes was
then once more stressed in review \cite{MSTV90}. Reference  \cite{SVSWW97} 
was the first that quoted the previous
result \cite{FRS76}. It 
incorporated most important in-medium effects in the cooling code, among them
nPBF and pPBF as equally 
important processes. The latter process was then
rediscovered in \cite{YKL99}
as negligible one 
due to ignorance there of correlations in weak interaction vertices.
Work 
\cite{P98} supported conclusion of \cite{SVSWW97} on importance of NPBF
processes as governing the cooling scenario at $T<T_c$.

The medium 
modifications of all the above mentioned rates result in a pronounced density
dependence (for NPBF processes mainly via dependence of the pairing gaps on the
density and dependence on the $NN$ correlation factors), 
which links the cooling behavior of a neutron
star decisively to its mass \cite{VS84,VS86,MSTV90,SVSWW97}. 
As the result, the above mentioned medium modifications lead to a
more rapid cooling than obtained in the {\em{"standard scenario"}}. Hence they
provide a possible explanation for the observed deviations of some of
the pulsar temperatures from the {\em{"standard"}} cooling.
Particularly, they provide a smooth transition from {\em{"standard"}}  to
{\em{"non-standard"}} cooling for increasing central star densities, i.e., star
masses. Thus by means of taking into account
of most important in-medium effects in the reaction rates 
one is indeed
able to achieve an appropriate
agreement with both the high as well as the low observed
pulsar temperatures that leads to the
new   {\em{"nuclear medium cooling scenario"}}.
Using a collection of modern EQS for nuclear matter, 
which covered both relativistic as well as
non-relativistic models,  \cite{SVSWW97} also has
demonstrated a relative robustness of these in-medium
cooling mechanisms against variations in the EQS of
dense NS matter (for EQS that allow for a wide dense hadronic region in NS).  

At initial stage ($T>T_{opac}$) the newly formed hot NS 
is opaque for neutrinos/antineutrinos. Within FOPE  model
the value $T_{opac}$ was estimated in \cite{FM79}. Elastic scatterings
were included in  \cite{SS79,S80} 
and pion condensation effect on the opacity was
discussed in  \cite{SS77}.
Medium effects dramatically affect the neutrino/antineutrino mean
free paths, since $\lambda_{\nu (\bar{\nu})} 
\propto 1/\mid M\mid^2$, where $M$ is the reaction matrix element.
Thereby, one-nucleon elastic scattering processes, like $N\nu\rightarrow N\nu$,
for energy and momentum transfer $\omega <  qv_{FN}$ 
are suppressed by $NN$ correlations \cite{VS87,MSTV90,RP97,PRPL99}.
Neutrino/antineutrino absorption in two-nucleon MMU and MNB
processes is substantially increased
with the density (since $\mid M\mid^2$ for MMU and MNB processes 
increases with the density) \cite{VS86,VS87,MSTV90}.
Thus more massive NS are opaque for neutrinos 
up to lower temperatures that also results in a delay of  
neutrino pulse. Within the QPA 
for the nucleons the value $T_{opac}$ was estimated
with taking into account of medium effects in  \cite{VS86,MSTV90}.
References \cite{VSKH87,HKSV88,MSTV90} considered possible consequences of
such a delay for supernova explosions. On the other hand, at $T>(1 \div 2)$MeV
one should take care of the neutrino/antineutrino
radiation in multiple $NN$ scatterings 
(Landau--Pomeranchuk--Migdal (LPM) effect) 
when averaged neutrino--antineutrino 
energy, 
$\omega_{\nu\bar{\nu}} \sim$ several $ T$,
becomes to be smaller than the nucleon width $\Gamma_N$ \cite{KV95}. Numerical
evaluations of $\Gamma_N$ in application to MNB processes
were done in \cite{SD99} and \cite{Y00} 
within the Br\"uckner scheme and the Bethe--Salpeter equation,
respectively. 
The LPM effect suppresses the
rates of the neutrino elastic scattering processes on nucleons
and also it suppresses
MNB rates.  
For NS of rather low mass ($\lsim 
M_{\bigodot}$) the suppression of the rates of neutral current processes
due to the multiple
collision coherence effect prevails over 
the enhancement due to the pion softening, 
and for sufficiently massive NS
($\gsim 
1.4 M_{\bigodot}$) the enhancement prevails the suppression. 
MMU emissivity remains to be 
almost unaffected by the LPM effect since averaged 
$\bar{\nu}e$ energy $\sim p_{Fe}$ is rather large ($\gg \Gamma_N$). 

Besides pion and kaon condensates, under current discussion is the
possibility of the presence of the quark matter at sufficiently high baryon
densities. Thereby, these phases may exist in most
massive NS. Also a particular discussion is devoted to the possibility of the
existence of less massive, stable or metastable, dense nuclei-stars glued by
the condensates or by strange quarks, cf. \cite{M71,VSC77,MSTV90,G96,W84}.
Uncertainty in these predictions relate to the poorly known EQS
at high baryon densities.
Several possible types of
stars differing in their values of the bag constant $B$ were discussed:
ordinary NS without any quark core, hybrid neutron stars (HNS)
with quark matter present only in their
deep interiors (for somewhat intermediate values of $B$), NS with large quark
core (QCNS), a narrow hadron layer  
and a crust typical for a NS, and quark stars (QS) with a tiny
crust of normal matter and with no crust (both for low $B$). 
If these objects are  produced in supernova explosions
within ordinary mechanism of blowing off the mantle, 
they are rather massive HNS. 
If an extra support for the blowing off the matter arises,
there may appear less massive objects like QCNS, QS and may be even  
objects of arbitrary size.

Especial interest for the discussion of cooling of HNS, QCNS, and QS 
is motivated
by recent works \cite{ARW98,RSSV98} which demonstrated possibility of
diquark condensates characterized by large pairing gaps ($\Delta_q \lsim
100~$MeV) in dense quark matter.
The two-flavor (2SC) and the three-flavor (3SC) color superconducting
phases allow for unpaired quarks of one color whereas in 
the color-flavor locking
(CFL) phase all the quarks are paired.  
Presence of large quark gap may significantly
affect the cooling history of HS, QCNS and QS. Thus it is interesting to
confront this possibility to the observational data.   
Cooling of QCNS, and QS
was discussed in \cite{BKV00,BGV00} and the case of HNS was considered
in \cite{PPLS00,BGV00}. In difference with 
\cite{BKV00,PPLS00} more recent work   
\cite{BGV00} incorporated the heat transport being important during a longer
period of time compared to that can be in absence of the color superconducting
quark matter.

The paper is organized as follows.  Sect. \ref{Nuclear} 
discusses basic ideas of
the Fermi liquid approach to description of nuclear
matter. The $NN$ interaction amplitude
is constructed with an explicit treatment of long-ranged soft pion mode 
and vertex renormalizations due to $NN$ correlations. The meaning of
the pion 
softening effect is clarified and a comparison of
MOPE  and FOPE models
is given. Also renormalization of the 
weak interaction in NS matter 
is performed. 
Sect. \ref{Neutrino} discusses the 
cooling of NS  at $T<T_{opac}$. 
Comparison of emissivities of MMU and MU processes  
shows a significant enhancement of in-medium rates.
Then we discuss DU-like processes  and  demonstrate medium effect due 
to vertex renormalizations. The role of
in-medium mechanisms in the 
cooling evolution of NS is then demonstrated
within a realistic cooling code.
Next we consider 
influence of in-medium effects on the neutrino mean free path
at initial stage of NS cooling. Essential role of multiple $NN$
collisions is discussed.  
Sect. \ref{How}
presents OTF in non-equilibrium closed diagram technique
in the framework of QPA
for the nucleons and also beyond the 
QPA incorporating genuine particle width effects. 
In the last Sect. \ref{Quark} following \cite{BGV00} 
we review recent results on the cooling of HNS.

\section{Nuclear Fermi liquid description}
\label{Nuclear}

\subsection{$NN$ interaction. Separation of  hard and soft modes}

At temperatures of our interest ($T\ll \varepsilon_{Fn}$) neutrons are
only slightly excited above their Fermi sea and all the processes occur in a
narrow vicinity of $\varepsilon_{Fn}$. In such a situation Fermi 
liquid approach
seems to be the most efficient one. 
Within this approach the long-ranged diagrams are treated explicitly
whereas short-scale diagrams are supposed to be the local quantities given by
phenomenological so called Landau-Migdal (LM) parameters.
Thus using argumentation of Fermi liquid theory \cite{L56,M67,M78,MSTV90}
the retarded $NN$ interaction amplitude is presented as follows 
(see also \cite{V93})
\begin{eqnarray}\label{NN-ampl}
\setlength{\unitlength}{1mm}
\parbox{10mm}{\begin{fmfgraph}(10,10)
\fmfpen{thick}
\fmfleftn{l}{2}
\fmfrightn{r}{2}
\fmfpolyn{full}{P}{4}
\fmf{fermion}{r1,P1}
\fmf{fermion}{P2,r2}
\fmf{fermion}{l2,P3}
\fmf{fermion}{P4,l1}
\end{fmfgraph}}
=
\setlength{\unitlength}{1mm}
\parbox{10mm}{\begin{fmfgraph}(10,10)
\fmfpen{thick}
\fmfleftn{l}{2}
\fmfrightn{r}{2}
\fmfpolyn{shaded,pull=1.4,smooth}{P}{4}
\fmf{fermion}{r1,P1}
\fmf{fermion}{P2,r2}
\fmf{fermion}{l2,P3}
\fmf{fermion}{P4,l1}
\end{fmfgraph}}
+
\setlength{\unitlength}{1mm}
\parbox{25mm}{\begin{fmfgraph}(30,10)
\fmfpen{thick}
\fmfleftn{l}{2}
\fmfrightn{r}{2}
\fmfpolyn{shaded,pull=1.4,smooth}{P}{4}
\fmfpolyn{full}{Pr}{4}
\fmf{fermion}{l2,P3}
\fmf{fermion}{P4,l1}
\fmf{fermion}{r1,Pr1}
\fmf{fermion}{Pr2,r2}
\fmf{fermion,left=.5,tension=.5}{Pr4,P1}
\fmf{fermion,left=.5,tension=.5}{P2,Pr3}
\end{fmfgraph}}
\,\,\,+
\setlength{\unitlength}{1mm}
\parbox{15mm}{\begin{fmfgraph}(35,10)
\fmfpen{thick}
\fmfleftn{l}{2}
\fmfrightn{r}{2}
\fmfpolyn{shaded,pull=1.4,smooth}{P}{4}
\fmfpolyn{full}{Pr}{4}
\fmf{fermion}{l2,P3}
\fmf{fermion}{P4,l1}
\fmf{fermion}{r1,Pr1}
\fmf{fermion}{Pr2,r2}
\fmf{fermion,left=.5,tension=.5}{Pr4,P1}
\fmf{heavy,left=.5,tension=.5}{P2,Pr3}
\end{fmfgraph}}
\end{eqnarray}
where
\begin{eqnarray}\label{irred}
\setlength{\unitlength}{1mm}
\parbox{10mm}{\begin{fmfgraph*}(10,10)
\fmfpen{thick}\fmfleftn{l}{2}
\fmfrightn{r}{2}
\fmfpolyn{shaded,pull=1.4,smooth}{P}{4}
\fmf{fermion}{r1,P1}
\fmf{fermion}{P2,r2}
\fmf{fermion}{l2,P3}
\fmf{fermion}{P4,l1}
\end{fmfgraph*}}
\,\,\,=\,\,\,
\parbox{10mm}{\begin{fmfgraph*}(10,10)
\fmfpen{thick}\fmfleftn{l}{2}
\fmfrightn{r}{2}
\fmfpolyn{hatched,pull=1.4,smooth}{P}{4}
\fmf{fermion}{r1,P1}
\fmf{fermion}{P2,r2}
\fmf{fermion}{l2,P3}
\fmf{fermion}{P4,l1}
\end{fmfgraph*}}
\,\,\,+\,\,
\parbox{25mm}{\begin{fmfgraph}(25,10)
\fmfpen{thick}\fmfleftn{l}{2}
\fmfrightn{r}{2}
\fmfpen{thick}
\fmf{fermion}{l2,ol}
\fmf{fermion}{ol,l1}
\fmf{fermion}{r1,o2}
\fmf{fermion}{o2,r2}
%
\fmf{dbl_wiggly,width=1thin}{ol,o2}
\end{fmfgraph}}\,\,\,\,\,\,.
\end{eqnarray}
The solid line presents the nucleon, whereas double-line, the
$\Delta$ isobar.
The double-wavy line corresponds to
the exchange of the free pion with inclusion of the contributions of the
residual 
$S$ wave $\pi NN$ interaction and $\pi\pi$ scattering, i.e. the residual
irreducible
interaction to the nucleon
particle-holes and delta-nucleon holes. The full particle-hole,
delta-nucleon hole and pion irreducible block (first block in (\ref{irred}))
is by its construction essentially more local then contributions given by 
explicitly presented graphs. Thereby, it is parameterized with the help of the
LM parameters. In the standard Landau Fermi liquid theory
fermions are supposed to be at their Fermi surface and
the Landau parameters are further expanded in Legendre polynomials in the 
angle between fermionic momenta.
Luckily, only zero and first harmonics enter physical quantities. 
The momentum dependence of the residual part of
nuclear forces is expected to be not as pronounced and 
one can avoid this expansion.
Then these parameters, i.e. 
$f_{nn}$, $f_{np}$ and $g_{nn}$, $g_{np}$ in scalar and 
spin channels respectively, are considered as slightly momentum 
dependent quantities. In principle, they should be calculated 
as functions of the density,
neutron and proton concentrations,
energy and momentum but, simplifying, one can extract them from analysis
of experimental data.

The part of interaction involving $\Delta$ isobar is  analogously constructed
\begin{equation} \label{gam1-d}
\setlength{\unitlength}{1mm}\parbox{10mm}{
\begin{fmfgraph}(10,10)
\fmfpen{thick}
\fmfset{arrow_len}{2mm}
\fmfset{arrow_ang}{30}
\fmfleftn{l}{2}
\fmfrightn{r}{2}
\fmfforce{(0.7w,0.3h)}{T1}
\fmfforce{(0.7w,0.7h)}{T2}
\fmfforce{(0.3w,0.7h)}{T3}
\fmfforce{(0.3w,0.3h)}{T4}
\fmfforce{(1.0w,0.0h)}{r1}
\fmfforce{(1.0w,1.0h)}{r2}
\fmfforce{(0.0w,1.0h)}{l2}
\fmfforce{(0.0w,0.0h)}{l1}
\fmfpolyn{shaded,smooth,pull=1.4}{T}{4}
\fmf{fermion}{r1,T1}
\fmf{heavy}{T2,r2}
\fmf{fermion}{l2,T3}
\fmf{fermion}{T4,l1}
\end{fmfgraph}}\,\,=\,\,
\setlength{\unitlength}{1mm}\parbox{10mm}{
\begin{fmfgraph*}(10,10)
\fmfpen{thick}\fmfset{arrow_len}{2mm}
\fmfset{arrow_ang}{30}
\fmfleftn{l}{2}
\fmfrightn{r}{2}
\fmfforce{(0.7w,0.3h)}{T1}
\fmfforce{(0.7w,0.7h)}{T2}
\fmfforce{(0.3w,0.7h)}{T3}
\fmfforce{(0.3w,0.3h)}{T4}
\fmfforce{(1.0w,0.0h)}{r1}
\fmfforce{(1.0w,1.0h)}{r2}
\fmfforce{(0.0w,1.0h)}{l2}
\fmfforce{(0.0w,0.0h)}{l1}
\fmfpolyn{hatched,smooth,pull=1.4}{T}{4}
\fmf{fermion}{r1,T1}
\fmf{heavy}{T2,r2}
\fmf{fermion}{l2,T3}
\fmf{fermion}{T4,l1}
\end{fmfgraph*}}
\,\,+\,\,
\setlength{\unitlength}{1mm}\parbox{20mm}{
\begin{fmfgraph*}(20,10)
\fmfpen{thick}\fmfset{arrow_len}{2mm}
\fmfset{arrow_ang}{30}
\fmfleftn{l}{2}
\fmfrightn{r}{2}
\fmfforce{(1.0w,0.0h)}{r1}
\fmfforce{(1.0w,1.0h)}{r2}
\fmfforce{(0.0w,1.0h)}{l2}
\fmfforce{(0.0w,0.0h)}{l1}
\fmf{fermion}{l1,ol}
\fmf{fermion}{ol,l2}
\fmf{fermion}{r1,or}\fmf{heavy}{or,r2}
\fmf{dbl_wiggly,width=thin}{ol,or}
\end{fmfgraph*}}\,\,\,\,\,\,\,\,.
\end{equation}
The main part of the $N\Delta$ interaction is due to the pion exchange.
Although information on local part of the $N\Delta$ interaction is rather
scarce, one can conclude \cite{MSTV90,SST99}
that the corresponding LM parameters
are essentially smaller then those for $NN$ interaction. 
Besides, at small transferred energies $\omega \ll m_{\pi}$ 
the $\Delta$--nucleon hole contribution is a smooth function of $\omega$
and $k$ in difference with the nucleon--nucleon hole ($NN^{-1}$)
contribution.
Therefore and also for simplicity we neglect the first graph in
r.h.s. of (\ref{gam1-d}).  

Straightforward resummation of (\ref{NN-ampl}) 
in neutral channel yields \cite{VS87,MSTV90}
\\
\\
\begin{equation}\label{neutr-land} 
\Gamma_{\alpha\beta}^R =\setlength{\unitlength}{1mm}
\parbox{10mm}{\begin{fmfgraph*}(10,10)
\fmfpen{thick}
\fmfleftn{l}{2}
\fmfrightn{r}{2}
\fmfpolyn{full}{P}{4}
\fmf{fermion}{r1,P1}
\fmf{fermion}{P2,r2}
\fmf{fermion}{l2,P3}
\fmf{fermion}{P4,l1}
\fmflabel{$\alpha$}{l1}\fmflabel{$\alpha$}{l2}
\fmflabel{$\beta$}{r1}\fmflabel{$\beta$}{r2}
\end{fmfgraph*}}
= C_0 \left({\cal{F}}_{\alpha\beta}^R +{\cal{Z}}_{\alpha\beta}^R 
{\vec{\sigma}}_1 
\cdot{\vec{\sigma}}_2 \right)+f_{\pi N}^2{\cal{T}}_{\alpha\beta}^R (
{\vec{\sigma}}_1\cdot{\vec{k}})
({\vec{\sigma}}_2\cdot{\vec{k}}),
\end{equation}
\begin{eqnarray}\label{c-f1}
{\cal{F}}_{\alpha\beta}^R &=&f_{\alpha\beta}\gamma (f_{\alpha\beta}),\,\,\,\, 
{\cal{Z}}_{nn}^R =g_{nn}\gamma (g_{nn}),\,\,\,\,
{\cal{Z}}_{np}^R =g_{np}\gamma (g_{nn}),\,\,\,\alpha , \beta =(n,p),
\nonumber \\
{\cal{T}}_{nn}^R &=&\gamma^2 (g_{nn})D^{R}_{\pi^0},\,\,\,\,
{\cal{T}}_{np}^R =-\gamma_{pp}\gamma (g_{nn})D^{R}_{\pi^0},\,\,\,\,
{\cal{T}}_{pp}^R =\gamma^2_{pp}D^{R}_{\pi^0},\nonumber \\
\gamma^{-1} (x)&=& 1-2 x C_0 A_{nn}^R ,\,\,\,\,\gamma_{pp}=(1-4 g
C_0  A_{nn}^R )
\gamma(g_{nn}),
\end{eqnarray}
$f_{nn}=f_{pp}=f+f'$, $f_{np}=f-f'$, $g_{nn}=g_{pp}=g+g'$, and
$g_{np}=g-g'$, dimensional normalization  factor 
is usually taken to be 
$C_0 =\pi^2
/[m_N p_F (\varrho_0 )]\simeq 300$~MeV$\cdot$ fm$^3$ 
$\simeq 0.77 m_{\pi}^{-2}$, 
$D^{R}_{\pi^0}$ is the full retarded Green function of $\pi^0$, 
$A_{\alpha\beta}$ 
is the corresponding $NN^{-1}$ loop (without spin degeneracy factor 2)
\\ \\
\begin{equation}\label{loo}
A_{\alpha\beta}=\,\,\,
\parbox{10mm}{
\begin{fmfgraph*}(20,10)\setlength{\unitlength}{1mm}
\fmfpen{thick}
\fmfleft{l}
\fmfright{r}
\fmf{fermion,left=.5,label=$\beta$,label.side=left,left=.8,tension=.8}{l,r}
\fmf{fermion,left=.5,label=$\alpha^{-1}$,
label.side=left,left=.8,tension=.8}{r,l}
\,\,
\end{fmfgraph*}},
A_{nn}(\omega \simeq q)\simeq
m^{*2}_n  (4\pi^2 )^{-1}
\left( \mbox{ln}\frac{1+v_{Fn}}{1-v_{Fn}} -2v_{Fn}\right) ,
\end{equation}
\\
\\
$A_{nn}\simeq -m^*_n p_{Fn}(2\pi^2)^{-1}$, for $\omega \ll
qv_{Fn}, q\ll 2p_{Fn}$,
and we for simplicity 
neglect proton hole contributions due to
a small concentration  of protons.
Resummation of (\ref{NN-ampl}) in the charged channel yields \\ 
\begin{equation}\label{ch-land} 
\widetilde{\Gamma}_{np}^R =\setlength{\unitlength}{1mm}
\parbox{10mm}{\begin{fmfgraph*}(10,10)
\fmfpen{thick}
\fmfleftn{l}{2}
\fmfrightn{r}{2}
\fmfpolyn{full}{P}{4}
\fmf{fermion}{r1,P1}
\fmf{fermion}{P2,r2}
\fmf{fermion}{l2,P3}
\fmf{fermion}{P4,l1}
\fmflabel{$p$}{l1}\fmflabel{$n$}{l2}
\fmflabel{$p$}{r1}\fmflabel{$n$}{r2}
\end{fmfgraph*}}
= C_0 \left(\widetilde{{\cal{F}}}_{np}^R +\widetilde{{\cal{Z}}}_{np}^R 
{\vec{\sigma}}_1 
\cdot{\vec{\sigma}}_2 \right)+f_{\pi N}^2\widetilde{{\cal{T}}}_{np}^R (
{\vec{\sigma}}_1\cdot{\vec{k}})
({\vec{\sigma}}_2\cdot{\vec{k}})\,,
\end{equation}
\begin{eqnarray}\label{c-f2}
\widetilde{{\cal{F}}}_{np}^R &=&2f' \widetilde{\gamma} (f' ),\,\,\, 
\widetilde{{\cal{Z}}}_{np}^R =2g' \widetilde{\gamma} (g' ),\,\,\,
\widetilde{{\cal{T}}}_{np}^R =\widetilde{\gamma}^2 (g' )
D^{R}_{\pi^- },\,\,\,\\
\widetilde{\gamma}^{-1} (x)&=& 1-4 x C_0 A_{np}^R \, .\nonumber
\end{eqnarray}
The LM parameters are rather unknown for isospin
asymmetric nuclear matter and for $\varrho >\varrho_0$. Although some 
evaluations of these quantities 
have been done, much work is still needed to get convincing results. 
Therefore for estimates we
will use the values extracted from atomic nucleus experiments. 
Using argumentation of a relative locality of these quantities  
we will suppose the LM 
parameters to be independent on the density for
$\varrho >\varrho_0$. One then can expect that the most uncertain 
will be the value
of the scalar constant $f$ due to essential role of the medium-heavy
$\sigma$ meson in this channel. But this parameter does not enter
the tensor force channel being most important in our case.
Unfortunately, there are also essential uncertainties in numerical values
of some of the  LM 
parameters even for atomic nuclei. 
These uncertainties are, mainly, due to attempts to
get the best fit of
experimental data in each concrete case slightly modifying 
parameterization used
for the residual part of the $NN$ interaction.
E.g., calculations \cite{M67} gave $f\simeq 0.25$, $f^{\prime}
\simeq 1$, $g\simeq 0.5$, $g^{\prime} \simeq 1$ whereas 
\cite{ST98,FZ95,BTF84}, including QP renormalization pre-factors,
derived the values $f\simeq 0$, $f^{\prime}
\simeq 0.5 \div 0.6$, $g\simeq 0.05\pm 0.1$, $g^{\prime} \simeq 1.1 \pm 0.1$.

Typical energies and momenta entering $NN$ interaction of our interest
are $\omega\simeq 0$ and $k\simeq p_{Fn}$. Then rough estimation yields
$\gamma (g_{nn} ,\omega\simeq 0 ,k\simeq p_{Fn}, \varrho =\varrho_0 )
\simeq 0.35 \div
0.45$.
For $\omega =k \simeq T$ typical for the weak 
processes with participation of
$\nu\bar{\nu}$ one has $\gamma^{-1} (g_{nn} ,\omega\simeq k\simeq T, \varrho
=\varrho_0 )\simeq 0.8 \div 0.9$.

\subsection{Virtual Pion Mode }
 
Resummation of diagrams yields the following Dyson equation for pions
\begin{equation} \label{pion-l}
\parbox{10mm}{\begin{fmfgraph}(20,30)
\fmfleft{l}
\fmfright{r}
\fmf{boson,width=thick}{l,r}\fmfforce{(0.0w,0.5h)}{l}
\fmfforce{(1.0w,0.5h)}{r}
\end{fmfgraph}}\,\,=\,\,\,\,
\parbox{10mm}{\begin{fmfgraph}(20,30)
\fmfleft{l}
\fmfright{r}
\fmf{boson,width=thin}{l,r}\fmfforce{(0.0w,0.5h)}{l}
\fmfforce{(1.0w,0.5h)}{r}\end{fmfgraph}}\,\,+\,\,\,\,
\parbox{25mm}{\begin{fmfgraph}(65,30)
\fmfleft{l}
\fmfright{r}
\fmf{boson}{l,ol}
\fmf{boson,width=thick}{or,r}
\fmfpoly{full}{or,pru,prd}
\fmfforce{(0.2w,0.5h)}{ol}
\fmfforce{(0.8w,0.5h)}{or}
\fmfforce{(0.6w,0.3h)}{prd}
\fmfforce{(0.6w,0.7h)}{pru}
\fmf{fermion,left=.5,tension=.5,width=1thick}{ol,pru}
\fmf{fermion,left=.5,tension=.5,width=1thick}{prd,ol}\fmfforce{(0.0w,0.5h)}{l}
\fmfforce{(0.9w,0.5h)}{r}
\end{fmfgraph}}+\,\,\,
\parbox{25mm}{\begin{fmfgraph}(60,30)
\fmfleft{l}
\fmfright{r}
\fmf{boson}{l,ol}
\fmf{boson,width=thick}{or,r}
\fmfforce{(0.2w,0.5h)}{ol}
\fmfforce{(0.8w,0.5h)}{or}\fmfforce{(1.0w,0.5h)}{r}\fmfforce{(0.0w,0.5h)}{l}
\fmf{heavy,left=.6,tension=.5}{ol,or}
\fmf{fermion,left=.6,tension=.5,width=1thick}{or,ol}
\fmfv{decor.shape=circle,decor.filled=full,decor.size=4thick}{or}\fmfforce{(0.0w,0.5h)}{l}
\fmfforce{(1.0w,0.5h)}{r}
\end{fmfgraph}}
+\,\,
\parbox{40mm}{\begin{fmfgraph*}(60,20)
\fmfleft{l}
\fmfright{r}
\fmf{boson}{l,ol}
\fmf{boson,width=thick}{or,r}
\fmfpoly{empty,label=$\Pi_{\rm res}^R$}{or,pru,plu,ol}
\fmfforce{(0.0w,0.0h)}{l}
\fmfforce{(1.0w,0.0h)}{r}
\fmfforce{(0.3w,0.0h)}{ol}
\fmfforce{(0.7w,0.0h)}{or}
\fmfforce{(0.7w,0.9h)}{pru}
\fmfforce{(0.3w,0.9h)}{plu}
\end{fmfgraph*}}
\end{equation}
The $\pi N\Delta$ full-dot-vertex includes a background
correction due to the presence of the higher resonances, 
$\Pi_{res}^{R}$ is the residual
retarded pion self-energy that includes the contribution of all the
diagrams which are not presented explicitly in (\ref{pion-l}), as $S$ wave 
$\pi NN$ and $\pi\pi$ scatterings (included by double-wavy line in 
(\ref{irred})).
The full vertex takes into account $NN$  
correlations  
\begin{equation} \label{pion-vert}
\parbox{20mm}{\begin{fmfgraph}(40,30)
\fmfleftn{l}{2}
\fmfright{r}
\fmfforce{(0.0w,0.0h)}{l1}
\fmfforce{(0.0w,1.0h)}{l2}
\fmfforce{(1.0w,0.5h)}{r}
\fmfpoly{full}{pr,pl2,pl1}
\fmfforce{(0.7w,0.5h)}{pr}
\fmfforce{(0.3w,0.7h)}{pl2}
\fmfforce{(0.3w,0.3h)}{pl1}
\fmf{fermion}{l2,pl2}
\fmf{fermion}{pl1,l1}
\fmf{boson}{pr,r}
\end{fmfgraph}}\,\,=
\,\,
\parbox{20mm}{\begin{fmfgraph}(40,30)
\fmfleftn{l}{2}
\fmfright{r}
\fmfforce{(0.0w,0.0h)}{l1}
\fmfforce{(0.0w,1.0h)}{l2}
\fmfforce{(1.0w,0.5h)}{r}
\fmf{fermion}{l2,ol,l1}
\fmf{boson}{ol,r}
\end{fmfgraph}}
\,\,+\,\,
\parbox{28mm}{\begin{fmfgraph}(60,30)
\fmfleftn{l}{2}
\fmfright{r}
\fmfforce{(0.0w,0.0h)}{l1}
\fmfforce{(0.0w,1.0h)}{l2}
\fmfforce{(1.0w,0.5h)}{r}
\fmfpoly{full}{pr,pl2,pl1}
\fmfforce{(0.8w,0.5h)}{pr}
\fmfforce{(0.6w,0.7h)}{pl2}
\fmfforce{(0.6w,0.3h)}{pl1}
\fmfpen{thick}\fmfpoly{hatched,smooth}{brd,bru,blu,bld}
\fmfforce{(0.2w,0.2h)}{bld}
\fmfforce{(0.2w,0.8h)}{blu}
\fmfforce{(0.4w,0.8h)}{bru}
\fmfforce{(0.4w,0.2h)}{brd}
\fmf{fermion}{l2,blu}
\fmf{fermion}{bld,l1}
\fmf{fermion,left=0.2}{bru,pl2}
\fmf{fermion,left=0.2}{pl1,brd}
\fmf{boson,width=1thin}{pr,r}
\end{fmfgraph}}
.
\end{equation}
Due to that the nucleon particle-hole part of $\Pi_{\pi^0}$
is $\propto \gamma (g_{nn})$ and the  
nucleon particle-hole part of $\Pi_{\pi^\pm}$
is $\propto \gamma (g' )$.  
The value of the $NN$ interaction in the pion channel is determined by the 
full pion propagator at small $\omega$ and 
$k \simeq p_{Fn}$, i.e. by the quantity 
$\widetilde{\omega}^2 (k)=-(D^R_{\pi})^{-1}(\omega =0, k,\mu_\pi )$. Typical 
momenta of our interest  are 
$ k\simeq p_{Fn}$. Indeed the momenta entering the $NN$ interaction 
in MU and MMU processes are $ k= p_{Fn}$, the momenta governing the MNB
are $ k=k_0$ \cite{VS86} where
the value $ k=k_0 \simeq (0.9\div1)p_{Fn}$
corresponds to the minimum of $\widetilde{\omega}^2 (k)$.
The quantity
$\widetilde{\omega}\equiv \widetilde{\omega}(k_0 )$ 
has the meaning of the {\em{effective pion gap}}. It is
different for $\pi^0$ and for $\pi^{\pm}$ since
neutral and charged channels are characterized by different diagrams
permitted by charge conservation, thus also 
depending on the value of the pion chemical
potential, $\mu_{\pi^{+}}\neq \mu_{\pi^{-}}\neq 0$, $\mu_{\pi^{0}}=0$. 
For $T\ll
\varepsilon_{Fn},\varepsilon_{Fp}$, one has   
$\mu_{\pi^{-}}=\mu_e =\varepsilon_{Fn}-\varepsilon_{Fp}$, 
as follows from equilibrium conditions for the reactions $n\rightarrow
p\pi^-$ and $n \rightarrow pe\bar{\nu}$. 

Change of the sign of  $\widetilde{\omega}^2$ symbolizes the pion condensate
phase transition.
Typical density behaviour of $\widetilde{\omega}^2$
is shown in Fig. 3.
\begin{figure}[h]
\centerline{
\includegraphics[height=4.5cm,clip=true]{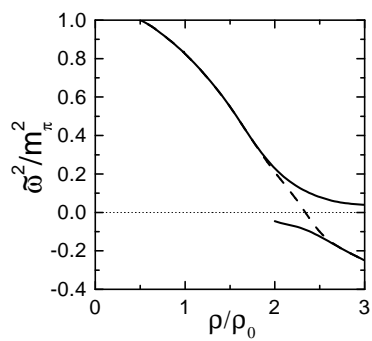}}
\caption{Effective pion gap (for $\mu_{\pi^{0}}=0$)  
versus baryon density, see  \cite{MSTV90}.}
\end{figure}
At $\varrho <(0.5\div0.7)\varrho_0$, 
one 
has $\widetilde{\omega}^2 =m^2_{\pi}-\mu_{\pi}^2$. 
For such  densities the value 
$\widetilde{\omega}^2 (p_{Fn})$ essentially deviates from 
$m^2_{\pi}-\mu_{\pi}^2$
tending to $m^2_{\pi}+p_{Fn}^2 -\mu_{\pi}^2$ 
in small density limit. 

At the critical point of the pion condensation 
($\varrho =\varrho_{c\pi}$)
the value $\widetilde{\omega}^2$ with artificially
switched off $\pi\pi$ fluctuations (dashed line in Fig. 3)
changes its sign. 
In reality $\pi\pi$ fluctuations are significant in the
vicinity of the critical point and there occurs the 
first-order phase transition to the
inhomogeneous  pion-condensate state \cite{D75,VM81,D82}. 
Thereby there are two branches (solid curves in Fig. 3) with positive and
respectively negative values for $\widetilde{\omega}^2 $.
Calculations of 
\cite{D82} demonstrated that at 
$\varrho >\varrho_{c\pi}$ the free energy of the state with 
$\widetilde{\omega}^2 >0$, where the pion mean field is zero, 
becomes larger than that of the
corresponding state with $\widetilde{\omega}^2 <0$ and a finite
mean field. Therefore at 
$\varrho >\varrho_{c\pi}$ the state with $\widetilde{\omega}^2 >0$ is
metastable and the state with $\widetilde{\omega}^2 <0$ and 
the pion mean field $\varphi_\pi \neq 0$ becomes the ground state.  

The quantity $\widetilde{\omega}^2 $ demonstrates
how much 
the virtual (particle-hole) mode with pion quantum numbers is softened
at given density.  The ratio
$\alpha =D_\pi [\mbox{med.}]/D_\pi [\mbox{vac.}]\simeq 6$
for $\varrho =\varrho_0$, $\omega =0$, $k=p_{FN}$
and for isospin symmetric nuclear matter.
However this essential so called {\em{"pion softening"}} \cite{M78}
does not significantly enhance the $NN$ scattering cross section due to a
simultaneous essential 
suppression of the $\pi NN$ vertex by $NN$ correlations. 
Indeed, the ratio of the $NN$ cross sections calculated with FOPE
and MOPE is
\begin{equation}
R=\frac{\sigma [\mbox{MOPE]}}{\sigma [\mbox{FOPE}]}
\simeq \frac{\gamma^4 (g' ,\omega\simeq 0,k\simeq p_{FN} )
(m^2_{\pi}+p_{FN}^2)^2}{\widetilde{\omega}^4 (p_{FN} )}\,,
\end{equation} 
and for $\varrho =\varrho_0$ we have $R\lsim 1$, 
whereas for $\varrho =2\varrho_0$ we already get
$R\sim 10$. 

As follows from numerical estimates of different $\gamma$ factors entering
(\ref{neutr-land}) and (\ref{ch-land}),
the main contribution to $NN$ interaction for $\varrho >\varrho_0$
is given by MOPE
\begin{equation}
\parbox{20mm}{\begin{fmfgraph}(20,20)
\fmfleftn{l}{2}
\fmfrightn{r}{2}
\fmfpoly{full}{p1,p2,p3,p4}
\fmfforce{(0.0w,0.0h)}{l1}
\fmfforce{(0.0w,1.0h)}{l2}
\fmfforce{(1.0w,0.0h)}{r1}
\fmfforce{(1.0w,1.0h)}{r2}
\fmfforce{(0.8w,0.0h)}{p1}
\fmfforce{(0.8w,1.0h)}{p2}
\fmfforce{(0.2w,1.0h)}{p3}
\fmfforce{(0.2w,0.0h)}{p4}
\fmf{fermion}{p4,l1}
\fmf{fermion}{l2,p3}
\fmf{fermion}{p2,r2}
\fmf{fermion}{r1,p1}
\end{fmfgraph}}\,\,\simeq\,\,\,\,\,\,\,\,\,\,\,\,\,\,
\parbox{20mm}{
\begin{fmfgraph}(20,40)
\fmfleftn{l}{2}
\fmfrightn{r}{2}
\fmfpoly{full}{ur,ul,uo}
\fmfpoly{full}{dr,do,dl}
\fmfforce{(0.0w,0.0h)}{l1}
\fmfforce{(0.0w,1.0h)}{l2}
\fmfforce{(1.0w,0.0h)}{r1}
\fmfforce{(1.0w,1.0h)}{r2}
\fmfforce{(0.2w,0.8h)}{ul}
\fmfforce{(0.8w,0.8h)}{ur}
\fmfforce{(0.5w,0.7h)}{uo}
\fmfforce{(0.2w,0.2h)}{dl}
\fmfforce{(0.8w,0.2h)}{dr}
\fmfforce{(0.5w,0.3h)}{do}
\fmf{fermion}{l2,ul}
\fmf{fermion}{ur,r2}
\fmf{fermion}{dl,l1}
\fmf{fermion}{r1,dr}
\fmf{boson,width=1thick}{do,uo}
\end{fmfgraph}}\,\,
\end{equation}
whether this channel (${\cal{T}}\propto  (
{\vec{\sigma}}_1\cdot{\vec{k}})
({\vec{\sigma}}_2\cdot{\vec{k}})$) of the reaction
is not forbidden or suppressed by some specific reasons like
symmetry, small momentum transfer, etc. 

The $\varrho$ meson 
contribution to $NN$ interaction is partially included in
$g_{\alpha \beta}$, other part contributing to ${\cal{T}}$ and 
$\widetilde{{\cal{T}}}$ is minor ($\propto \widetilde{\omega}^2
/m_\varrho{^2}$).
Indeed, using that
$(\vec{\sigma}_1 \times \vec{k})
(\vec{\sigma}_2 \times\vec{k})=k^2 \vec{\sigma}_1\vec{\sigma}_2 -
(\vec{\sigma}_1  \vec{k})\cdot (\vec{\sigma}_2 \vec{k})
$
the $\rho$ exchange can be cast as
\begin{eqnarray}
\delta \Gamma^{R}_{N_1 N_2}=\left(\frac{f_\rho}{m_\rho}\right)^2
\left\{ \frac{k^2 (\vec{\sigma}_1\vec{\sigma}_2 )
\gamma^{2}}{[\omega^{2} -m_{\rho}^{2} -k^{2}-\Pi_\rho^R ]}-
\frac{(\vec{\sigma}_1\vec{k}) (\vec{\sigma}_2 \vec{k})
\gamma^{2}}{[\omega^{2} -m_{\rho}^{2} -k^{2}-\Pi_\rho^R ]}\right\}.
\end{eqnarray}
We may omit contribution $\mbox{Re}\Pi^{R}_\rho$ in the Green function
since $\mbox{Re}\Pi^{R}_\rho \ll m_{\rho}^{2}$ for 
$\rho \sim m_{\pi}^{3}$
under consideration; $\gamma$ factor is the same as for the corresponding
pion (charged or neutral).
First term is supposed to be included in phenomenological
value of the corresponding Landau--Migdal parameter leading to its 
momentum dependence. Due to that the value 
$g'$ gets a 30\% decrease (rather then an increase discussed in 
some works) with the momentum at $k=p_{FN}$ compared 
to the corresponding value at $k=0$, see \cite{PSTF95}. Thus one can think
that $g' (0)>g' (p_{FN})$.
Second term can be dropped as small ($\lsim 1/30$ of
MOPE contribution at $\rho =\rho_0$).

Another worry \cite{BFS93,BBLW95} 
was expressed in connection with 
quasielastic polarization transfer experiment at LAMPF, EMC experiment and the
Drell--Yan experiment at Fermi Lab which did not observe an
expected in several works a pronounced
pion excess in nuclei. 
First optimistic estimates demonstrated
the pion excess to be at (15-30)\% level that 
was ruled out by different models which analyzed mentioned 
experimental results including artificially enhanced pion contribution
and suppressing other important contributions. 
However it may well be that only in a very narrow 
vicinity of  $\varrho_{c\pi}$ (far beyond $\varrho_0$)
pion fluctuations grow at $T=0$ and only
at
$T\neq 0$ pion excess must essentially grow
in substantially wider density interval \cite{VM81,D82}.

Let us
show that some effect should be although it may be numerically rather small.
Let estimate a contribution of the virtual pion sea (region of 
small energies
$\omega < kv_F$) to the sum rule. 
It  is given by
\begin{eqnarray}\label{sumcon}
S(kv_F )=\int_{0}^{k v_F} 2\omega A_{\pi} \frac{d\omega}{ 2\pi}
=\frac{2}{\pi \beta}\left[ k v_F -\frac{\widetilde{\omega}^2 (k)}
{\beta}
\mbox{arctan}\left(
\frac{\beta k v_F}{\widetilde{\omega}^2 (k)}\right)\right] \ll 1,
\end{eqnarray}
$A_{\pi}$ is the pion spectral function,
$S(\infty ) =1,\,\,\,\, \beta =\frac{m_{N}^{\ast 2}kf_{\pi N}^2}{\pi}$, 
$m^*_N$ is the nonrelativistic effective mass given by the relation
$(d\epsilon (p)/d~p)|_{p_F}=p_F /m^*_N$.
The quantity $S(kv_F )$ having no singularity even at
$\widetilde{\omega}^2 (k_0 )\rightarrow 0$ is, thus, rather
insensitive to the pion softening effect. 
Therefore mentioned experimental
analysis  might be
not as critical to the
value $\widetilde{\omega}^2 (k_0 )$.

The physical reason of a 
contribution of virtual pions 
at $\omega < kv_F$ is clear. It is the well 
known Landau damping associated with possibility of the virtual pion decay
to the nucleon particle and hole, i.e. just with the Pauli blocking. 
As known from $\pi N$ scattering experiments, 
pions interact with nucleons. Thus must be a contribution of virtual pions
to the sum-rule for pion spectral function.
There is no other way out. 
Therefore on a question "Where virtual pions are?" we would still
suggest an old fashion answer: "They are 
hidden inside the matter". The question which
remains is only "What is the actual
value of the pion excess?" To properly answer it
one needs besides the pionic contribution
to elaborate all relevant specific contributions related
to the phenomenon analyzed in the given concrete experiment. E.g., 
quasielastic polarization transfer data are reasonably fitted with a
slightly increased value of the Landau--Migdal parameter $g'$ 
and with inclusion of $S$ wave repulsion at small energies
\cite{MSTV90,E94,DE94},
EMC experiment
is reproduced with inclusion of the soft pion
mode and other relevant effects \cite{MOF96}.

Thus instead of FOPE$+\varrho$ exchange, 
as the model of $NN$ interaction which has been used in
\cite{FM79}
in calculation of the emissivities of the two-nucleon reactions within
the {\em{"standard scenario"}} of NS cooling,
one should use the full $NN$ interaction given by 
(\ref{neutr-land}) and (\ref{ch-land})
or, simplifying, approximated by  its MOPE part.

\subsection{Renormalization of the weak interaction.}\label{ren}

The full weak coupling vertex that takes into account $NN$  
correlations is determined by  (\ref{pion-vert}) where now the wavy line
should be replaced by the lepton pair.
Thus for the vertex of our interest, $N_1 \rightarrow N_2 l \bar{\nu}$,
we obtain \cite{VS87,MSTV90}
\begin{eqnarray}\label{betav}
V_\beta =\frac{G}{\sqrt{2}}\left[ \widetilde{\gamma}(f' )l_0 -g_A
\widetilde{\gamma}(g' )
\vec{l}\vec{\sigma} \right],
\end{eqnarray}
for the $\beta$ decay 
and
\begin{eqnarray}\label{nn-cor}
V_{nn} =-\frac{G}{2\sqrt{2}}\left[ \gamma (f_{nn} )l_0 -g_A\gamma (g_{nn} )
\vec{l}\vec{\sigma} \right],\,\,\,
V_{pp}^N =\frac{G}{2\sqrt{2}}\left[ \kappa_{pp} l_0 -g_A\gamma_{pp}
\vec{l}\vec{\sigma} \right],
\end{eqnarray}
\begin{eqnarray}\label{pp-cor}
\kappa_{pp}=c_V -2f_{np}\gamma (f_{nn})C_0 A_{nn},\,\,
\gamma_{pp}=\left( 1 -4g C_0 A_{nn} \right) \gamma (g_{nn}),
\end{eqnarray}
for processes on the neutral currents $N_1 N_2 \rightarrow N_1 N_2 \nu
\bar{\nu}$, $V_{pp}=V_{pp}^N +V_{pp}^{\gamma}$,
$G\simeq 1.17\cdot 10^{-5}$~GeV$^{-2}$ is the Fermi weak coupling 
constant, $c_V =1-4\sin^2 \theta_W$, $\sin^2 \theta_W \simeq 0.23$,
$g_A\simeq 1.26$ is the axial-vector coupling constant,
and $l_\mu =\bar{u}(q_1 )\gamma_\mu (1-\gamma_5 )u(q_2 )$ is the lepton 
current. The pion contribution $\sim \vec{q}^{\,2}$ is small for
typical $\mid\vec{q}\mid \simeq T$ or $p_{Fe}$, and for simplicity 
is omitted. In medium the value of $g_A$ (i.e. $g_A^*$) slightly
decreases with the density that can be easily incorporated, cf. the
Brown--Rho scaling idea \cite{BR96}. Please notice that 
with a decrease of $g_A^*$, in particular,
the value $\varrho_{c\pi}$ increases remaining however to be finite ($\simeq
2\varrho_{0}$
according to \cite{BCDM75} where this decrease of $g_A^*$ up to 1 
was discussed)
due to attractive
contribution of the $\Delta$ isobar in (\ref{pion-l}), 
whereas one would
expect $\varrho_{c\pi}\rightarrow \infty$ for $g_A^*\rightarrow 1$ ignoring
$\Delta$ contribution.

The $\gamma$ factors renormalize the corresponding vacuum vertices.
These factors are essentially different for different processes involved.
The matrix elements of the neutrino/antineutrino scattering processes
$N\nu\rightarrow N\nu$ and of MNB 
behave differently in dependence on the energy-momentum  transfer and 
whether $N=n$
or $N=p$ in the weak coupling vertex. Vertices 
\begin{eqnarray}\label{nu-scat}
\,\,\,\,\,\,\,\,\,\,\,\,\,\,\,\,\,\,\,\,\,\,\,\,
\parbox{10mm}{\begin{fmfgraph*}(50,35)
\fmfpen{thick}
\fmfleft{l1,l2}
\fmfright{r1,r2}
\fmf{fermion}{l1,T2}
\fmf{fermion}{T3,r1}
\fmf{fermion,width=1thin}{l2,o3,r2}
\fmf{dashes,width=1thin}{T1,o3}
\fmfpoly{full}{T2,T3,T1}
\fmflabel{$\nu$}{l2}
\fmflabel{$\nu$}{r2}\fmflabel{$N$}{l1}\fmflabel{$N$}{r1}\end{fmfgraph*}}
\,\,\,\,\,\,\,\,\,\,\,\,\,\,\,\,\,\,\,\,\,\,\,\, ,
\,\,\,\,\,\,\,\,\,\,\,\,\,\,\,\,\,\,\,\,\,\,\,\,
\parbox{10mm}{\begin{fmfgraph*}(50,35)
\fmfpen{thick}
\fmfleft{l1}
\fmfright{r1,r2,r3}
\fmf{fermion}{l1,T2}
\fmf{fermion}{T3,r1}
\fmf{fermion,width=1thin}{o3,r3}\fmf{fermion,width=1thin}{r2,o3}
\fmf{dashes,width=1thin}{T1,o3}
\fmfpoly{full}{T2,T3,T1}
\fmflabel{$\nu$}{r3}
\fmflabel{$\bar{\nu}$}{r2}\fmflabel{$N$}{l1}\fmflabel{$N$}{r1}\end{fmfgraph*}}
\,\,\,
\end{eqnarray}\\
are modified by the correlation factors (\ref{c-f1})
and (\ref{c-f2}). For $N=n$ these are
$\gamma (g_{nn}, \omega ,q)$ and 
$\gamma (f_{nn}, \omega ,q)$ leading to an 
enhancement  of the cross sections for $\omega >
qv_{Fn}$ and to a suppression for $\omega < qv_{Fn}$.
Renormalization of the proton vertex (vector part of 
$V_{pp}^N +V_{pp}^{\gamma}$) is governed by the processes \cite{VS87,VKK98}
\begin{eqnarray}\label{p-vert}
\,\,\,\,\,\,\,\,\,\,\,\,\,\,\,\,
\parbox{30mm}{\begin{fmfgraph*}(100,60)
\fmfpen{thick}
\fmfleft{l1}
\fmfright{r1,r2,r3}
\fmf{fermion,label=$p$,label.side=right}{l1,T2}
\fmf{fermion,label=$p$,label.side=right}{T3,r1}
\fmf{fermion,width=1thin}{o3,r3}\fmf{fermion,width=1thin}{r2,o3}
\fmf{fermion,label=$n$,label.side=left,left=.8,tension=.8}{T1,o2}
\fmf{fermion,label=$n^{-1}$,label.side=left,left=.8,tension=.8}{o2,T4}
\fmf{dashes,width=1thin}{o2,o3}
\fmfpolyn{full,smooth,pull=1.4}{T}{4}
\fmflabel{$\nu$}{r3}
\fmflabel{$\bar{\nu}$}{r2}\end{fmfgraph*}}
\,\,\,\,\,\,\,\,\,\,\,\,\,\,\,\,\,\,\,\,+
\,\,\,\,
\parbox{30mm}{\begin{fmfgraph*}(100,80)
\fmfpen{thick}
\fmfleft{l1}
\fmfright{r1,r2,r3}
\fmf{fermion,label=$p$,label.side=right}{l1,T2}
\fmf{fermion,label=$p$,label.side=right}{T2,r1}
\fmf{fermion,width=1thin}{o3,r3}\fmf{fermion,width=1thin}{r2,o3}
\fmf{fermion,width=1thin,label=$e$,label.side=left,left=.8,tension=.8}{T1,o2}
\fmf{fermion,width=1thin,
label=$e^{-1}$,label.side=left,left=.8,tension=.8}{o2,T1}
\fmf{dashes,width=1thin}{o2,o3}\fmf{photon,width=1thick,
label=$\gamma_m$,label.side=left}{T2,T1}
\fmflabel{$\nu$}{r3}
\fmflabel{$\bar{\nu}$}{r2}\end{fmfgraph*}}\,\,\,\,\,\,\,\,\,
\,\,\,\,\,\,\,\,\,+\,... 
\end{eqnarray}
being forbidden in vacuum. For the systems with $1S_0$ proton--proton
pairing, $\propto g_A^2$ 
contribution to the squared matrix element (see (\ref{nn-cor}))
is compensated by the corresponding contribution of the diagram with anomalous
Green functions of protons. The vector current term is $\propto c_V^2$
in vacuum whereas it is $\propto \kappa_{pp}^2$ in medium (by first
graph (\ref{p-vert})).
Thereby the corresponding vertices with participation of proton  
are enhanced  in medium compared
to small vacuum value ($\propto c_V^2 \simeq 0.006$)
leading to enhancement of the cross sections, up to 
$\sim 10\div10^2$ times for $1.5\div3\varrho_0$ depending on 
parameter choice. It does not contradict to the statement of
ref. \cite{VS87} that correlations are rather 
suppressed in the weak interaction
vertices 
at $\varrho \leq \varrho_0$. Enhancement with the density
comes from estimation (\ref{loo}) that directly 
follows from equation (2.30) of \cite{VS87}.
Also enhancement factor (up to $\sim 10^2$)  
comes from the virtual 
in-medium photon ($\gamma_m$)
whose propagator contains $1/m_{\gamma}^2
\sim 1/e^2$, where $m_{\gamma}$ is the effective spectrum gap,
that compensates small
$e^2$ factor from electromagnetic vertices, see \cite{VKK98}. It is
included by replacement $V_{pp}^N \rightarrow V_{pp}$. 
Other processes permitted in
intermediate states like processes with $pp^{-1}$ and with the pion are
suppressed
by a small proton density and by 
$q^2\sim T^2$ pre-factors, respectively. 
First diagram (\ref{p-vert})
was considered in \cite{VS87}, where the
pPBF process was suggested, and then in \cite{MSTV90,SVSWW97}, and 
it was shown that nPBF and pPBF processes
may give contributions of the same order of magnitude. 
Finally, with electron and nucleon
correlations included we recover this statement and numerical
estimate \cite{VS87,SVSWW97}
(in spite of mentioned misprints in the result \cite{VS87}).
Several
subsequent papers \cite{YKL99,KHY99,YLS99}
rediscovered pPBF process
ignoring
nucleon and electron correlation effects. 
(Although the authors 
were personally informed, they continue to insist \cite{YKGH00}
on incorrect treatment, giving priority to  their result.)
Contribution of second diagram for pPFB process was 
recently incorporated in \cite{L00}. 

Paper \cite{KV99} gives another example demonstrating that, although 
the vacuum branching ratio of the kaon decays is
$\Gamma(K^-\rightarrow e^-+\nu_e)/\Gamma(K^-\rightarrow
\mu^-+\nu_\mu)\approx 2.5\times 10^{-5}$,
in medium (due to $\Lambda p^{-1}$ decays of virtual $K^-$) it becomes to be
of order of unit.
Thus we again see that in dependence of what reaction channel is considered
in-medium effects may or strongly enhance the reaction rates 
or substantially suppress them. Ignorance of these
effects may lead to misleading results where one struggling for numerical
factors $\sim 1$ may loose by orders of magnitude larger factors.

\subsection{Inconsistencies of FOPE model}

Since FOPE model became the base of the {\em{"standard scenario"}} for
cooling simulations we would like first to 
demonstrate principal inconsistencies of the model for the
description of interactions in
dense ($\varrho \gsim \varrho_0$) baryon medium. 
The only diagram in FOPE model which contributes to the MU and NB is
\begin{equation}
\parbox{20mm}{\begin{fmfgraph*}(100,50)
\fmfleft{l1,l2}
\fmfright{r1,r2,r3,r4}
\fmf{fermion}{l1,o1,r1}
\fmf{fermion}{l2,o2}\fmf{fermion,label=$f_{\pi N}$,label.side=left}{o2,o3}
\fmf{fermion}{o3,r2}
\fmf{fermion}{r3,o4}\fmf{fermion}{o4,r4}
\fmf{dots,width=1thin}{o1,o2}\fmf{dashes,width=1thin}{o3,o4}
\fmflabel{$f_{\pi N}$}{o1}
\end{fmfgraph*}}
\end{equation}
Dots symbolize FOPE. This is first available Born approximation diagram, i.e.
second order perturbative contribution in $f_{\pi N}$.
In order to be theoretically consistent 
one should use perturbation theory up to  the very same second order in
$f_{\pi N}$ for all the quantities. E.g., pion spectrum is then determined by
pion polarization operator expanded up to the very same order in $f_{\pi N}$
\begin{equation}\label{pio}
\omega^2 \simeq m_{\pi}^2 +k^2 +\Pi^R_0 (\omega , k, \varrho ), \,\,\,\,\,
\Pi^R_0 (\omega , k, \varrho )=
\,\,\,\,\,\,\,\,\,\,
\parbox{50mm}{
\begin{fmfgraph*}(50,20)
\fmfleft{l}
\fmfright{r}
\fmf{fermion,left=.8,tension=.8}{o1,o2}
\fmf{fermion,left=.8,tension=.8}{o2,o1}\fmf{dots,
label=$f_{\pi N}$,label.side=left}{l,o1}
\fmf{dots,label=$f_{\pi N}$,label.side=right}{o2,r}
\fmfforce{(0.0w,0.5h)}{l}
\fmfforce{(0.2w,0.5h)}{o1}
\fmfforce{(0.8w,0.5h)}{o2}
\fmfforce{(1w,0.5h)}{r}
\end{fmfgraph*}}
\end{equation}
The value $\Pi^0 (\omega , k, \varrho )$ is easily calculated containing no
any uncertain parameters. For $\omega \rightarrow 0$, 
$k\simeq p_F$ of our
interest and for isospin symmetric matter 
\begin{equation}
\Pi^R_0 
\simeq -\alpha_0
-\I \beta_0 \omega ,\,\,\alpha_0 \simeq \frac{2m_N p_F k^2 f_{\pi N}^2}
{\pi^2}>0, \,\,\beta_0 \simeq\frac{m_N^2 k f_{\pi N}^2}{\pi}>0.
\end{equation}
Replacing this value to  (\ref{pio}) we obtain a solution with
$\I \omega  <0$ already for $\varrho >0.3 \varrho_0$ that 
would mean appearance of the 
pion condensation. Indeed, the mean field 
begins to increase with the time passage 
$\varphi \sim \mbox{exp}({-\I \omega t}) \sim \mbox{exp}({\alpha t/\beta})$ 
until repulsive 
$\pi\pi$ interaction will not stop its growth. 
But it is experimentally proven that there is no pion
condensation in atomic nuclei, i.e. even at  $\varrho =\varrho_0$.
The puzzle is solved as follows.
FOPE model does not work for such densities. One should replace FOPE by the
full $NN$ interaction given by (\ref{neutr-land}), (\ref{ch-land}).
Essential part of this interaction is due to MOPE with vertices corrected
by $NN$ correlations.
Also the $NN^{-1}$ part of the
pion polarization operator is corrected by $NN$ correlations. Thus
\begin{equation}
\setlength{\unitlength}{1mm}\parbox{20mm}{\begin{fmfgraph}(20,10)
\fmfleft{l}
\fmfright{r}
\fmf{boson}{l,ol}
\fmf{boson,width=thin}{or,r}
\fmfpoly{full}{or,pru,prd}
\fmfforce{(0.2w,0.5h)}{ol}
\fmfforce{(0.8w,0.5h)}{or}
\fmfforce{(0.6w,0.3h)}{prd}
\fmfforce{(0.6w,0.7h)}{pru}
\fmf{fermion,left=.5,tension=.5,width=1thick}{ol,pru}
\fmf{fermion,left=.5,tension=.5,width=1thick}{prd,ol}
\end{fmfgraph}}\,\,\,\simeq \,\,\, \Pi^R_0 (\omega , k, \varrho )\gamma (g' ,
\omega , k, \varrho ) 
\end{equation}
being suppressed by the factor $\gamma (g' ,\omega =0, k\simeq p_F , \varrho
\simeq \varrho_0 )\simeq 0.35 \div 0.45$. 
Final solution of the dispersion relation
(\ref{pio}), now with full 
$\Pi$ instead of $\Pi^0$, yields $\I\omega >0$ for $\varrho =\varrho_0$
whereas the solution with  $\I \omega <0$, which shows the
beginning of
pion condensation, appears only for 
$\varrho > \varrho_{c\pi}>\varrho_0$.

\section{Neutrino cooling of neutron stars 
}\label{Neutrino}
\subsection{Emissivity of MMU process}

Since DU process is forbidden up to sufficiently high density $\varrho_{cU}$,
the main contribution for $\varrho <\varrho_{cU}$ and $T_{opac}>T>T_c$
comes from MMU processes schematically
presented by two diagrams of Fig. 1.
MNB reactions give smaller contribution \cite{VS86}.
For densities $\varrho \ll \varrho_0$ the main part of the $NN$
interaction amplitude is given by the residual $NN$ interaction.
In this case the $NN$ interaction amplitude can be better
treated within the $T$ matrix approach which sums up the ladder diagrams in the
particle-particle channel rather then by LM parameters. 
Calculations of MNB processes with the
vacuum $T$ matrix \cite{HPR00}
found essentially smaller emissivity then that given by 
FOPE. Also  the in-medium
modifications of the $T$ matrix additionally suppress the rates of both
MMU and MNB processes, see \cite{BRSSV95}. Thus even at
small densities the FOPE model may give only a rough 
estimate of the emissivity of two nucleon processes. 
At $\varrho\gsim (0.5\div0.7)\, \varrho_0$ reactions in 
particle-hole channel and more specifically with participation of the soft 
pion mode  begin to dominate. 

Evaluations \cite{VS86,VSKH87,HKSV88,MSTV90}
showed that the dominating contribution to MMU rate 
comes from the
second diagram of Fig. \ref{fig:graph1}, namely from 
contributions to it given by the  first two diagrams
of the series 
\\
\begin{equation}\label{MMU-diag}
\parbox{30mm}{
\begin{fmfgraph*}(60,80)
\fmfleftn{l}{2}
\fmfrightn{r}{4}
\fmfpoly{full}{ur,ul,uo}
\fmfpoly{full}{dr,do,dl}\fmfpen{thick}
\fmfforce{(0.0w,0.0h)}{l1}
\fmfforce{(0.0w,1.0h)}{l2}
\fmfforce{(1.0w,0.0h)}{r1}
\fmfforce{(1.0w,1.0h)}{r2}\fmfforce{(1.2w,0.3h)}{r3}\fmfforce{(1.2w,0.7h)}{r4}
\fmfforce{(0.8w,0.5h)}{o2}\fmfforce{(.5w,0.5h)}{o1}
\fmfforce{(0.4w,0.8h)}{ul}
\fmfforce{(0.6w,0.8h)}{ur}
\fmfforce{(0.5w,0.7h)}{uo}
\fmfforce{(0.4w,0.2h)}{dl}
\fmfforce{(0.6w,0.2h)}{dr}
\fmfforce{(0.5w,0.3h)}{do}
\fmf{fermion}{l2,ul}
\fmf{fermion}{ur,r2}
\fmf{fermion}{l1,dl}
\fmf{fermion}{dr,r1}
\fmf{boson,width=1thick,label=$\pi^0$}{do,o1}
\fmf{boson,width=1thick,label=$\pi^+$}{o1,uo}
\fmf{fermion,width=1thin}{r3,o2}
\fmf{fermion,width=1thin}{o2,r4}\fmf{dashes,width=1thin}{o1,o2}
\fmflabel{$\bar{\nu}$}{r3}\fmflabel{$e$}{r4}\fmflabel{$p$}{r2}
\fmflabel{$n$}{r1}\fmflabel{$n$}{l1}\fmflabel{$n$}{l2}
\end{fmfgraph*}}\,+\,\,\,\,\,\,\,\,\,
\parbox{30mm}{
\begin{fmfgraph*}(60,80)
\fmfleftn{l}{2}
\fmfrightn{r}{4}
\fmfpoly{full}{ur,ul,uo}
\fmfpoly{full}{dr,do,dl}\fmfpen{thick}
\fmfforce{(0.0w,0.0h)}{l1}
\fmfforce{(0.0w,1.0h)}{l2}
\fmfforce{(1.0w,0.0h)}{r1}
\fmfforce{(1.0w,1.0h)}{r2}\fmfforce{(1.2w,0.3h)}{r3}\fmfforce{(1.2w,0.7h)}{r4}
\fmfforce{(0.9w,0.5h)}{o2}\fmfforce{(0.65w,0.5h)}{o1}
\fmfforce{(0.4w,0.9h)}{ul}
\fmfforce{(0.6w,0.9h)}{ur}\fmfforce{(0.5w,0.8h)}{uo}\fmfforce{(0.5w,0.7h)}{o4}
\fmfforce{(0.4w,0.1h)}{dl}
\fmfforce{(0.6w,0.1h)}{dr}
\fmfforce{(0.5w,0.2h)}{do}\fmfforce{(0.5w,0.3h)}{o3}
\fmf{fermion}{l2,ul}
\fmf{fermion}{ur,r2}
\fmf{fermion}{l1,dl}
\fmf{fermion}{dr,r1}
\fmf{boson,width=1thick,label=$\pi^0$}{do,o3}
\fmf{boson,width=1thick,label=$\pi^+$}{o4,uo}
\fmf{fermion,width=1thin}{r3,o2}
\fmf{fermion,width=1thin}{o2,r4}\fmf{dashes,width=1thin}{o1,o2}
\fmflabel{$\bar{\nu}$}{r3}\fmflabel{$e$}{r4}\fmflabel{$p$}{r2}
\fmflabel{$n$}{r1}\fmflabel{$n$}{l1}\fmflabel{$n$}{l2}
\fmfpoly{full}{o1,o5,o6}
\fmfforce{(0.6w,0.6h)}{o5}\fmfforce{(0.6w,0.4h)}{o6}
\fmfpoly{full}{or,pru,prd}
\fmfforce{(0.5w,0.3h)}{or}
\fmfforce{(0.6w,0.4h)}{prd}
\fmfforce{(0.4w,0.4h)}{pru}
\fmfpoly{full}{o4,pru1,prd1}
\fmfforce{(0.5w,0.7h)}{o4}
\fmfforce{(0.6w,0.6h)}{prd1}
\fmfforce{(0.4w,0.6h)}{pru1}
\fmf{fermion,left=.5,tension=.5,width=1thick}
{prd1,prd}
\fmf{fermion,left=.5,tension=.5,width=1thick,label=$n^{-1}$,label.side=left}
{pru,pru1}
\end{fmfgraph*}}\,+\,\,\,\,\,\,\,\,\,
\parbox{30mm}{
\begin{fmfgraph*}(60,60)
\fmfleftn{l}{2}
\fmfrightn{r}{4}
\fmfpoly{full}{ur,ul,uo}\fmfpen{thick}
\fmfpoly{full}{dr,do,dl}
\fmfforce{(0.0w,0.2h)}{l1}
\fmfforce{(0.0w,0.8h)}{l2}
\fmfforce{(1.2w,0.2h)}{r1}
\fmfforce{(1.2w,0.8h)}{r2}\fmfforce{(1.2w,1.1h)}{r3}\fmfforce{(1.2w,1.3h)}{r4}
\fmfforce{(0.9w,1.1h)}{o2}\fmfforce{(0.9w,0.9h)}{o1}
\fmfforce{(0.4w,0.8h)}{ul}
\fmfforce{(0.6w,0.8h)}{ur}
\fmfforce{(0.5w,0.7h)}{uo}
\fmfforce{(0.4w,0.2h)}{dl}
\fmfforce{(0.6w,0.2h)}{dr}
\fmfforce{(0.5w,0.3h)}{do}
\fmf{fermion,width=1thick}{l2,ul}
\fmf{fermion,width=1thick,label=$n$,label.side=left}{ur,ddl}
\fmf{fermion,width=1thick,label=$p$,label.side=right}{ddr,r2}
\fmf{fermion,width=1thick}{l1,dl}
\fmf{fermion,width=1thick}{dr,r1}
\fmf{boson,width=1thick,label=$\pi^0$}{do,uo}
\fmf{fermion,width=1thin}{r3,o2}
\fmf{fermion,width=1thin,label=$e$,label.side=left}{o2,r4}
\fmf{dashes,width=1thin}{o1,o2}
\fmflabel{$\bar{\nu}$}{r3}
\fmflabel{$n$}{r1}\fmflabel{$n$}{l1}\fmflabel{$n$}{l2}
\fmfpoly{full}{ddr,o1,ddl}\fmfforce{(0.8w,0.8h)}{ddl}
\fmfforce{(1.0w,0.8h)}{ddr}
\end{fmfgraph*}}\,+\,\,...\,\,\,\,\,\,\,\,\,
\end{equation}
whereas the third diagram, which naturally generalizes the corresponding
MU(FOPE) contribution,
gives only a small correction for $\varrho \gsim \varrho_0$.
The emissivity from the two first diagrams 
in a simplified notation \cite{MSTV90,SVSWW97} reads
\begin{eqnarray}\label{GU}
  &\varepsilon&^{MMU} [\mbox{MOPE}]\simeq 2.4\cdot 10^{24} ~ T^{8}_9
\left(\frac{\varrho
    }{\varrho_0}\right)^{10/3} \frac{(m^{\ast}_n )^3 m^{\ast}_p }{m_N^4}  
\left[\frac{m_\pi}{ \alpha \tilde{\omega}_{\pi^0}(p_{Fn})}\right]^4
\nonumber \\
  &\times&
\left[\frac{m_\pi}{ \alpha \tilde{\omega}_{\pi^\pm}(p_{Fn})}\right]^4\, 
\Gamma^8 ~ F_1 
  \zeta(\Delta_n)~\zeta(\Delta_p)~ {\rm\frac{erg}{cm^3 ~sec}} ~,
\end{eqnarray} 
where  
$T_9 =T/10^9$ K is the temperature, 
$m^*_n$ and $m^*_p$ are the nonrelativistic
effective neutron and proton masses, and the correlation 
factor $\Gamma^8$ is roughly
\begin{eqnarray}\label{gam-bet}
&\Gamma&^8 \simeq
\gamma_\beta^2 (\omega\simeq p_{Fe}, q\simeq p_{Fe})
\gamma^2 (g_{nn} , \omega \simeq 0, k=p_{Fn})\widetilde{\gamma}^4 
(g' , 0, p_{Fn}), \nonumber \\
&\gamma&^2_{\beta}(\omega , q)=\frac{\widetilde{\gamma}^2 (f^{\prime},
\omega , q)+3g^2_{A}
\widetilde{\gamma}^2 (g^{\prime}, \omega , q)}{1+3g^2_{A}},
\end{eqnarray}
and the second term in the factor 
\begin{equation}
F_1
\simeq 1+\frac{3}{4\widetilde{\gamma}^2 (g' , 0, p_{Fn})
\gamma_\beta^2 (\omega\simeq p_{Fe}, q\simeq p_{Fe})}\,
\left(\frac{\varrho}{\varrho_0}\right)^{2/3} 
\end{equation}
is the
contribution of the pion decay from intermediate states (first diagram
(\ref{MMU-diag})). 
The quantity $\Gamma$ effectively 
accounts for an appropriate product of the $NN$ correlation factors
in different $\pi N_1 N_2$ vertices.  For charged pions the value
$\mu_\pi \neq 0$ is incorporated in the expression for
the effective pion gap, for neutral pions $\mu_\pi = 0$.
The value $\alpha \sim 1$ depends on condensate structure,
$\alpha =1$ for
$\varrho<\varrho_{c\pi}$, and $\alpha =\sqrt{2}$ 
taking
account of the new excitations on the ground of the charged
$\pi$ condensate vacuum for $\varrho >\varrho_{c\pi}$.  
The factor 
\begin{eqnarray}
  \zeta(\Delta_{N}) \simeq \left\{ \begin{array}{ll} {\rm
    exp}\left(-\Delta_N/T\right) & \,\,\,\mbox{ $ T \le T_{cN}$},\\ 1 &
  \,\,\,\mbox{$T > T_{cN}$}, \,\,\, N=(n,p)
\end{array}
\right.
\end{eqnarray} 
estimates the suppression caused by the $nn$ and $pp$
pairings. Deviation of these factors from
simple exponents can be incorporated as in  \cite{YLS99}.

The ratio of the emissivities of MMU(MOPE) and MU(FOPE)
is roughly 
\begin{equation}\label{VS-FM}
\frac{\varepsilon^{MMU}[\mbox{MOPE}]}{\varepsilon^{MU}[\mbox{FOPE}]} 
\simeq 10^{3}
\frac{
\gamma^{2}(g_{nn}, 0, p_{Fn})\widetilde{\gamma}^{2}(g^{\prime}, 0, p_{Fn})}{
\widetilde{\omega}^{4}_{\pi^0}(p_{Fn})\widetilde{\omega}^{4}_{\pi^\pm}(p_{Fn})}
\left(\varrho /\varrho_0 \right)^{10/3}.
\end{equation}
For $\varrho \simeq \varrho_0$ this ratio is
$\sim 10$ whereas being estimated with the only third diagram
(\ref{MMU-diag}) it would be less then unit.

\subsection{Emissivity of DU-like processes}

{\bf{NPBF processes}}.
The one-nucleon processes
with neutral currents given by the second diagram (\ref{nu-scat})
for $N =(n, p)$ 
are forbidden at $T>T_c$ by energy-momentum conservations but they can 
occur at $T<T_c$. Then physically the processes relate to NPBF, see Fig. 2.
However they need special techniques to be calculated \cite{FRS76,VS87}.
These processes $n\rightarrow n\nu\bar{\nu}$ and $p\rightarrow
p\nu\bar{\nu}$ play very important role in the cooling of
superfluid NS, see \cite{VS87,SV87,MSTV90,SVSWW97,P98,YLS99}. 
The emissivity
for 3 types of neutrinos is given by \cite{VS87,SV87}  
\begin{eqnarray}\label{neutr-nu}
&&\varepsilon [\mbox{nPBF}] =
\frac{3\cdot 4G^2 \left(\xi_1 \gamma^2 (f_{nn} ) +\xi_2 g_{A}^{*2} 
\gamma^2 (g_{nn})
\right)
p_{Fn} m_n^{\ast}\Delta_{n}^7}{15\pi^5} I\left(\frac{
\Delta_n }{T}\right)
\nonumber \\
&&\simeq \zeta \cdot 10^{28}\left( \frac{\varrho }{\varrho_0 }\right)^{1/3} 
\frac{m^{\ast}_n }{m_N}\left( \frac{\Delta_n }{\mbox{MeV}}\right)^7 
I\left(\frac{
\Delta_n }{T}\right)  \,
\frac{\mbox{erg}}{\mbox{cm}^3\cdot \mbox{sec}}, \,\,\,\, T<T_{cn},
\end{eqnarray}
where $\xi_1 =1$, $\xi_2 =0$ for $S$-pairing
and $\xi_1 =2/3$, $\xi_2 =4/3$ for $P$-pairing, compare with the
result \cite{YLS99} where $\xi_1 =1$, $\xi_2 =2$ were obtained. 
I 
removed some misprints existed in \cite{FRS76,VS87,SV87,MSTV90}. 
For neutrinos
$\omega (q) =q$ and correlations are not so essential as it would be  for 
$\omega
\ll q$. Taking $\gamma^2 \simeq 1.3$ in the range of $S$-pairing we get  $\zeta
\simeq 5$ whereas  for the
$P$-pairing with $g_A^{*}\simeq 1.1$
we obtain $\zeta
\simeq 9$,  in agreement with numerical evaluations \cite{VS87} ($\zeta
\simeq 7$) used then
within
the cooling code in \cite{SVSWW97},
$I(x)=\int_{0}^{\infty} \mbox{ch}^5 y dy /\left(\exp (x\mbox{ch}y
)+1\right)^2 $, 
$I(x\gg 1)\simeq \mbox{exp}(-2\Delta/T)\sqrt{\pi T/4\Delta}$,
that serves an  appropriate asymptotic (\ref{neutr-nu}). 
Emissivity of the process 
$p \rightarrow p\nu \bar{\nu}$ is given by \cite{VS87}
\begin{eqnarray}\label{prot-nu}
\varepsilon (\mbox{pPBF})&=&
\frac{12 G^2 (\xi_1 \kappa^2_{pp} +\xi_2 g_{A}^{*2} \gamma^2_{pp}+\xi_3 )
p_{Fp} m_p^{\ast}\Delta_{p}^7}{15\pi^5}I\left(\frac{\Delta_p }{T}\right) ,
\,
T<T_{cp},
\end{eqnarray}
and $\xi_2 =0$ for
protons paired in $S$-state in NS matter,
$\xi_3 \lsim 1$ is due to the
second diagram (\ref{p-vert}) and has a
complicated structure \cite{VKK98,L00}. For the process
(\ref{prot-nu})
the part of $NN$ and $ee$ correlations is especially important. One has
$\kappa_{pp}^2\sim 0.05\div 1$ for $1\div 3\varrho_0$ and $\xi_3
\sim 1$,
and $\kappa_{pp}^2+\xi_3 \sim 1$
instead of a small $c_V^{2} \simeq 0.006$ value
in absence of
correlations, see discussion in subsection \ref{ren}. 
Thereby, in agreement with
\cite{VS87,SVSWW97,VKK98,L00}, the emissivity of the process
$p\rightarrow p\nu\bar{\nu}$ can be compatible with that for 
$n \rightarrow n\nu\bar{\nu}$ in dependence on the relation between
$\Delta_p$ and $\Delta_n$.

NPBF processes are
very efficient  for $T<T_c$ competing with MMU processes. The former win
the content for not too massive stars. 
Analysis of above processes supports also
our general conclusion on the crucial role of in-medium effects in the 
cooling scenario.

{\bf{Pion (kaon) condensate processes}}.
The $P$ wave pion condensate can be of three
types: $\pi_s^+$, $\pi^\pm$, and $\pi^0$ with different values of the
critical densities $\varrho_{c\pi}=(\varrho_{c\pi^\pm}, \varrho_{c\pi^+_{s}}, 
\varrho_{c\pi^{0}}$), see \cite{M78}. Thus above the threshold density 
for the pion 
condensation of the
given type, the neutrino emissivity of the MMU
process (\ref{GU}) is to be 
supplemented by the corresponding PU processes\\
\\
\begin{equation}\label{picond}
\parbox{30mm}{
\begin{fmfgraph*}(60,50)
\fmfleft{l}
\fmfrightn{r}{3}
\fmfpoly{full}{ul,ur,uo}\fmfpen{thick}
\fmfpoly{full}{dl,dr,do}
\fmfforce{(0.0w,0.6h)}{l}
\fmfforce{(1.3w,0.6h)}{r1}
\fmfforce{(0.7w,0.9h)}{r2}\fmfforce{(0.7w,1.3h)}{r3}
\fmfforce{(0.95w,0.0h)}{o}
\fmfforce{(0.4w,0.9h)}{o2}
\fmfforce{(0.4w,0.7h)}{uo}
\fmfforce{(0.5w,0.6h)}{ur}
\fmfforce{(0.3w,0.6h)}{ul}
\fmfforce{(0.95w,0.6h)}{dl}
\fmfforce{(1.15w,0.6h)}{dr}
\fmfforce{(1.05w,0.5h)}{do}
\fmf{fermion,width=1thick}{l,ul}
\fmf{fermion,width=1thick,label=$p$}{ur,dl}
\fmf{fermion,width=1thick}{dr,r1}
\fmf{boson,width=1thick,label=$\pi^-_c $,label.side=left}{do,o}
\fmf{fermion,width=1thin,label=$\bar{\nu}$,label.side=right}{r3,o2}
\fmf{fermion,width=1thin,label=$e$}{o2,r2}\fmf{dashes,width=1thin}{uo,o2}
\fmfv{decor.shape=cross,decor.size=4thick}{o}
\fmflabel{$n$}{r1}\fmflabel{$n$}{l}
\end{fmfgraph*}}\,\,\,\,\,\,\,\,\,\,\,\,\,\, , \,\,\,
\parbox{30mm}{
\begin{fmfgraph*}(60,50)
\fmfleft{l}
\fmfrightn{r}{3}
\fmfpoly{full}{ul,ur,uo}\fmfpen{thick}
\fmfpoly{full}{dl,dr,do}
\fmfforce{(0.0w,0.6h)}{l}
\fmfforce{(1.3w,0.6h)}{r1}
\fmfforce{(0.7w,0.9h)}{r2}\fmfforce{(0.7w,1.3h)}{r3}
\fmfforce{(1.15w,0.0h)}{o}
\fmfforce{(0.4w,0.9h)}{o2}
\fmfforce{(0.4w,0.7h)}{uo}
\fmfforce{(0.5w,0.6h)}{ur}
\fmfforce{(0.3w,0.6h)}{ul}
\fmfforce{(0.95w,0.6h)}{dl}
\fmfforce{(1.15w,0.6h)}{dr}
\fmfforce{(1.05w,0.5h)}{do}
\fmf{fermion,width=1thick}{l,ul}
\fmf{fermion,width=1thick,label=$p$}{ur,dl}
\fmf{fermion,width=1thick}{dr,r1}
\fmf{boson,width=1thick,label=$\pi^+_c$,label.side=left}{do,o}
\fmf{fermion,width=1thin,label=$\bar{\nu}$,label.side=right}{r3,o2}
\fmf{fermion,width=1thin,label=$e$}{o2,r2}\fmf{dashes,width=1thin}{uo,o2}
\fmfv{decor.shape=cross,decor.size=4thick}{o}
\fmflabel{$n$}{r1}\fmflabel{$n$}{l}
\end{fmfgraph*}}\,\,\,\,\,\,\, , \,\,\,\,\,\,\,
\parbox{30mm}{
\begin{fmfgraph*}(60,50)
\fmfleft{l}
\fmfrightn{r}{3}
\fmfpoly{full}{ul,ur,uo}\fmfpen{thick}
\fmfpoly{full}{dl,dr,do}
\fmfforce{(0.0w,0.6h)}{l}
\fmfforce{(1.3w,0.6h)}{r1}
\fmfforce{(0.7w,0.9h)}{r2}\fmfforce{(0.7w,1.3h)}{r3}
\fmfforce{(1.05w,0.0h)}{o}
\fmfforce{(0.4w,0.9h)}{o2}
\fmfforce{(0.4w,0.7h)}{uo}
\fmfforce{(0.5w,0.6h)}{ur}
\fmfforce{(0.3w,0.6h)}{ul}
\fmfforce{(0.95w,0.6h)}{dl}
\fmfforce{(1.15w,0.6h)}{dr}
\fmfforce{(1.05w,0.5h)}{do}
\fmf{fermion,width=1thick}{l,ul}
\fmf{fermion,width=1thick,label=$n$}{ur,dl}
\fmf{fermion,width=1thick}{dr,r1}
\fmf{boson,width=1thick,label=$\pi^0_c $,label.side=left}{do,o}
\fmf{fermion,width=1thin,label=$\bar{\nu}$,label.side=right}{r3,o2}
\fmf{fermion,width=1thin,label=$\nu$}{o2,r2}\fmf{dashes,width=1thin}{uo,o2}
\fmfv{decor.shape=cross,decor.size=4thick}{o}
\fmflabel{$n$}{r1}\fmflabel{$n$}{l}
\end{fmfgraph*}}\,\,\,... \,\,\,\,\,\,\,\,\,\,
\end{equation}
The emissivity of the charged pion condensate processes  
with  inclusion of the
$NN$ correlation effect (in a simplified treatment) renders, see
\cite{VS84,MSTV90},
\begin{eqnarray}\label{cond}
  \varepsilon [\mbox{PU}] &\simeq &7\cdot 10^{26}~ \frac{p_{Fn}}{m_\pi }
  \frac{m^{\ast}_n m^{\ast}_p}{m_N^2}  
\gamma^2_\beta (p_{Fe}, p_{Fe})~
\widetilde{\gamma}^{2}(g^{\prime}, 0, p_{Fn})
  T^{6}_9 \mbox{sin}^2 \theta 
{\rm\frac{erg}{cm^3 sec}}.
\end{eqnarray}
Here $\varrho>\varrho_{c\pi}$ and
$\sin\theta \simeq
\sqrt{2|\tilde{\omega}^2|/m_\pi^2}$ for $\theta \ll 1$. 
Of the same order of
magnitude are emissivities of other possible $\pi$ condensate reactions, e.g.
for $n\pi_c^0\rightarrow npe\bar{\nu}$ process at $\theta \ll 1$
the numerical factor is about 
two times larger. Since $\pi^{\pm}$
condensation probably reduces the energy gaps of the superfluid states
by an order of magnitude, see 
\cite{T80}, we may assume that
superfluidity vanishes above $\varrho_{c\pi}$. 
Finally we note that though the PU
processes have genuinely one-nucleon phase-space volumes, their
contribution to the emissivity is suppressed relative to the
DU by an additional $\widetilde{\gamma}^{2}(g^{\prime}, 0, p_{Fn})$
suppression factor due to existence
of the extra ($\pi NN$) vertex in the former case.

\begin{figure}
\centering\psfig{figure=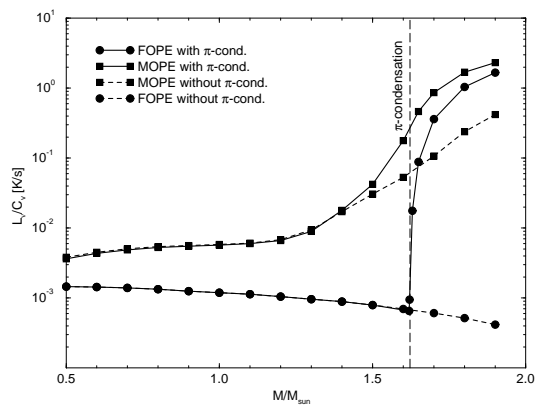,height=6.8cm,angle=-90}
\caption[]{Cooling rate due to neutrino emission as a function of 
  star mass for a representative temperature of $T=3\times
  10^8$~K. 
Superfluidity is neglected.  \label{fig:rate}}
\end{figure}

Fig. \ref{fig:rate} compares the mass dependence of the
neutrino cooling rates $L_\nu /C_V$, $L_\nu$ is the neutrino luminosity,
$C_V$ is the heat capacity,
for MMU(MOPE)
and MU(FOPE)
for non-superfluid matter. For the solid curves, the neutrino
emissivity in pion-condensed matter 
is taken into account according to (\ref{cond})
and (\ref{GU}) with the parameter $\alpha =\sqrt{2}$. To be conservative
we took 
$\varrho_{c\pi}\simeq 3\varrho_0$. Dashed
curves correspond to the model where no pion condensation is
allowed. As one sees, the medium polarization effects included in
MMU may result in three order of magnitude increase of the cooling
rate for the most massive stars. Even for stars of a rather low mass the
cooling rate of MMU is still several times larger than for MU
because even in this case the more efficient rate is given by the
reactions shown by the right diagram in Fig. 1 (first two diagrams
of (\ref{MMU-diag})).  The cooling rates for 
the NS of $M=1.8M_\odot$ with and without pion condensate differ
only moderately (by factor of 5 in this model). 
If we would take $\varrho_{c\pi}$ 
to be smaller the ratio of emissivity PU to MMU 
would decrease and could even 
become $\lsim 1$ in the vicinity of $\varrho_{c\pi}$.
Opposite, the reaction rates for the FOPE model
are rather independent of the star's mass for the stars with masses
below the critical value $1.63M_{\bigodot}$, at which 
transition into the pion condensed phase occurs, and then jump
to the typical PU value. 
It is to be stressed that  contrary to FOPE model, 
the MOPE model \cite{VS86} consistently takes
into account the pion softening effects for $\varrho <\varrho_{c\pi}$ 
and both the
pion condensation and pion softening effects
on the ground of the condensate for $\varrho >\varrho_{c\pi}$.
For $\varrho >\varrho_{cK}$ the kaon condensate
processes come into play. 
Most popular is the idea of the
$S$ wave $K^-$ condensation (e.g. see \cite{BKPP88}) which is allowed at 
$\mu_e >m^*_{K^-}$ due to possibility of the reaction $e\rightarrow
K^-\nu$. Analogous condition for the pions, $\mu_e >m^*_{\pi^-}$,
is not fulfilled owing to a strong $S$ wave $\pi NN$
repulsion \cite{M78,MSTV90} 
(again in-medium effect!) otherwise $S$ wave $\pi^-$ condensation would occur
at smaller densities then $K^-$ condensation.
The neutrino emissivity of the $K^-$ condensate processes  
is given by equation analogous to (\ref{cond})
with a different $NN$ correlation factor and an additional suppression factor
due to a small contribution of the Cabibbo angle. However 
qualitatively scenario
that permits kaon condensate processes is analogous to that with the 
pion condensate processes.

{\bf{Other resonance processes}}.
There are many other reaction channels allowed in the medium. E.g, any Fermi
liquid permits propagation of  
zero sound excitations of different symmetry related to the pion
and the quanta of a more  local interaction determined here via
$f_{\alpha ,\beta}$ and $g_{\alpha ,\beta}$. These
excitations being present at $T\neq 0$ may also participate in the 
neutrino reactions. The most essential contribution comes from the
neutral current processes \cite{VS86} given by first two diagrams of the
series\\
\begin{eqnarray}\label{res-pr}
\parbox{30mm}{
\begin{fmfgraph*}(60,60)
\fmfpen{thick}
\fmfforce{(0.7w,1.0h)}{r3}\fmfforce{(0.3w,1.0h)}{r4}
\fmfforce{(0.5w,0.8h)}{o2}\fmfforce{(0.5w,0.65h)}{o1}
\fmfforce{(0.0w,0.5h)}{uo}\fmfforce{(0.3w,0.5h)}{o4}
\fmfforce{(1.0w,0.5h)}{do}\fmfforce{(0.7w,0.5h)}{o3}
\fmf{dots,width=1thick}{do,o3}
\fmf{dots,width=1thick}{o4,uo}
\fmf{fermion,width=1thin}{o2,r3}
\fmf{fermion,width=1thin}{o2,r4}\fmf{dashes,width=1thin}{o1,o2}
\fmflabel{$\nu$}{r3}\fmflabel{$\bar{\nu}$}{r4}
\fmfpoly{full}{o1,o5,o6}
\fmfforce{(0.4w,0.6h)}{o5}\fmfforce{(0.6w,0.6h)}{o6}
\fmfpoly{full}{or,pru,prd}
\fmfforce{(0.7w,0.5h)}{or}
\fmfforce{(0.6w,0.6h)}{prd}
\fmfforce{(0.6w,0.4h)}{pru}
\fmfpoly{full}{o4,pru1,prd1}
\fmfforce{(0.3w,0.5h)}{o4}
\fmfforce{(0.4w,0.6h)}{prd1}
\fmfforce{(0.4w,0.4h)}{pru1}
\fmf{fermion,left=.5,tension=.5,width=1thick}
{prd1,prd}
\fmf{fermion,left=.5,tension=.5,width=1thick,label=$n^{-1}$,label.side=left}
{pru,pru1}
\end{fmfgraph*}}\,+\,\,\,\,\,\,\,\,\,\,
\parbox{30mm}{
\begin{fmfgraph*}(60,50)
\fmfleft{l}
\fmfrightn{r}{3}
\fmfpoly{full}{ul,ur,uo}\fmfpen{thick}
\fmfpoly{full}{dl,dr,do}
\fmfforce{(0.0w,0.8h)}{l}
\fmfforce{(1.3w,0.8h)}{r1}
\fmfforce{(0.7w,1.1h)}{r2}\fmfforce{(0.7w,1.3h)}{r3}
\fmfforce{(1.05w,0.2h)}{o}
\fmfforce{(0.4w,1.1h)}{o2}
\fmfforce{(0.4w,0.9h)}{uo}
\fmfforce{(0.5w,0.8h)}{ur}
\fmfforce{(0.3w,0.8h)}{ul}
\fmfforce{(0.95w,0.8h)}{dl}
\fmfforce{(1.15w,0.8h)}{dr}
\fmfforce{(1.05w,0.7h)}{do}
\fmf{fermion,width=1thick}{l,ul}
\fmf{fermion,width=1thick,label=$n$}{ur,dl}
\fmf{fermion,width=1thick}{dr,r1}
\fmf{dots,width=1thick}{do,o}
\fmf{fermion,width=1thin}{r3,o2}
\fmf{fermion,width=1thin,label=$\nu$}{o2,r2}\fmf{dashes,width=1thin}{uo,o2}
\fmflabel{$\bar{\nu}$}{r3}
\fmflabel{$n$}{r1}\fmflabel{$n$}{l}
\end{fmfgraph*}}\,\,\,\,\,\,\,\,+\,\,\, ...
\end{eqnarray}
Here the dotted line is zero sound quantum of appropriate symmetry. 
These are the resonance processes (second, of DU-type) 
analogous to those processes going on the condensates with the
only difference that the rates of reactions with
zero sounds are proportional to the thermal occupations
of the corresponding spectrum branches whereas the rates of the
condensate processes are proportional to the 
modulus squared of condensate mean field.
Contribution of the resonance
reactions is as a role 
rather small due to a small phase space volume ($q\sim T$)
associated with zero sounds. Please also bear in mind an analogy
of the processes (\ref{res-pr}) with the corresponding 
phonon processes in the crust.   

{\bf{DU processes}}.
The proper DU processes in matter, as $n\rightarrow pe^-\bar{\nu}_e$
and $ pe^-\rightarrow n\nu_e$,
\begin{eqnarray} \label{du}
\,\,\,\,\,\,\,\,\,\,\,\,\,\,\,\,\,\,\,\,\,\,\,\,
\parbox{10mm}{\begin{fmfgraph*}(50,35)
\fmfpen{thick}
\fmfleft{l1}
\fmfright{r1,r2,r3}
\fmf{fermion}{l1,T2}
\fmf{fermion}{T3,r1}
\fmf{fermion,width=1thin}{o3,r3}\fmf{fermion,width=1thin}{r2,o3}
\fmf{dashes,width=1thin}{T1,o3}
\fmfpoly{full}{T2,T3,T1}
\fmflabel{$e$}{r3}
\fmflabel{$\bar{\nu}$}{r2}\fmflabel{$n$}{l1}\fmflabel{$p$}{r1}\end{fmfgraph*}}
\,\,\,\,\,\,\,\,\,\,\,\,\,\,\,\,\,\,\,\,\,\,\,\,\,\,+
\,\,\,\,\,\,\,\,\,\,\,\,
\parbox{10mm}{\begin{fmfgraph*}(50,35)
\fmfpen{thick}
\fmfleft{l1,l2}
\fmfright{r1,r2}
\fmf{fermion}{l1,T2}
\fmf{fermion}{T3,r1}
\fmf{fermion,width=1thin}{l2,o3}\fmf{fermion,width=1thin}{o3,r2}
\fmf{dashes,width=1thin}{T1,o3}
\fmfpoly{full}{T2,T3,T1}
\fmflabel{$e$}{l2}
\fmflabel{$\nu$}{r2}\fmflabel{$p$}{l1}\fmflabel{$n$}{r1}\end{fmfgraph*}}
\,\,\,\,\,\,\,\,\,\,\,\,\,\,\,\,\,\,\,\,\,\,\,\,\,
\end{eqnarray}\\
should also be treated with the full vertices.  They are
forbidden up to the density $\varrho_{cU}$
when triangle inequality $p_{Fn}<p_{Fp}+p_{Fe}$
begins to fulfill. For traditional EQS like that given by the
variational theory \cite{APR98} DU processes are 
permitted for $\varrho >5\varrho_0$. The
emissivity of the DU processes 
renders
\begin{eqnarray}\label{neutr-DU}
\varepsilon^{DU} \simeq 1.2
\cdot 10^{27}\frac{m^{\ast}_{n}m^{\ast}_p}{m_N^2}
\left( \frac{\mu_l}{100 \mbox{MeV}} \right) \gamma^2_{\beta}\mbox{min}\left[
\zeta
(\Delta_n ) ,\zeta(\Delta_p )\right]
T_{9}^6 \,\,{\rm\frac{erg}{cm^3~sec}}, 
\end{eqnarray}
where $\mu_l =\mu_e =\mu_{\mu}$ is the chemical
potential of the leptons in MeV. In addition to the usually exploited
result \cite{LPPH91}, 
(\ref{neutr-DU}) contains $\gamma_\beta^2$ pre-factor (\ref{betav})
due to $NN$ correlations in
the $\beta$ decay vertices, see
\cite{VS87,SVSWW97}.  

It was realized in  \cite{VKZC00} 
that the softening of the pion mode in dense neutron matter 
could also give rise to a rearrangement 
of single-fermion degrees of freedom due to violation of Pomeranchuk
stability condition for $\varrho =\varrho_{cF}<\varrho_{c\pi}$.
It may result in organization of an extra Fermi sea for
$\varrho_{cF}<\varrho <\varrho_{c\pi}$ and at small momenta 
$p<0.2 p_{Fn}$,
that in its turn opens a DU channel of neutrino cooling of 
NS from the corresponding layer.
Due to the new feature of a temperature-dependent 
neutron effective mass, $m_n^{\ast} \propto 1/T$, we may anticipate extra 
essential
enhancement of the DU process, corresponding to a reduction 
in the power of the temperature dependence from $T^6$ to $T^5$.
At early hot stage this layer becomes to be opaque for the neutrinos
slowing the transport from the massive NS  core to the exterior.

Basing on the Brown--Rho scaling idea \cite{BR96}, 
we argued in \cite{V97} for the charged 
$\varrho$ meson 
condensation at a relevant density ($\varrho_{c\varrho}\sim 3
\varrho_0$ if $m^*_{\varrho}$ would decrease to that density up to  
$m_{\varrho}/2$). If happened, it would open DU reaction already
for $\varrho <
\varrho_{c\varrho}$  and close it for $\varrho >
\varrho_{c\varrho}$ due to an essential modification of the nuclear asymmetry
energy.

\subsection{Comparison with soft $X$ ray data}
\begin{figure*}
\centering\psfig{figure=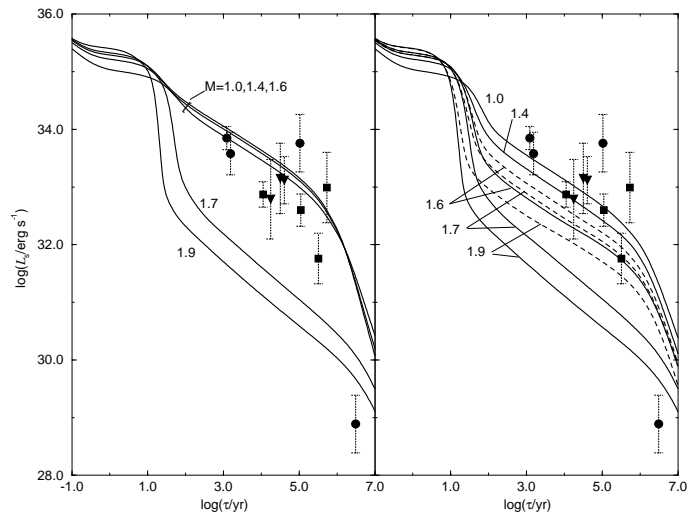,height=9cm,angle=-90}
\caption[]{Cooling of non-superfluid NS models of 
  different masses constructed for the HV EOS \cite{SVSWW97}. 
Two graphs refer to cooling via MU(FOPE) +PU (left) and MMU(MOPE)+PU (right). 
In both cases, pion condensation is taken into account for the solid curves at
$\varrho>\varrho_{c\pi}$, i.e. for $M>1.6 M_\odot $ ($\varrho_{c\pi}$ is
chosen to be $3\varrho_0$). The 
dashed curves in the right graph refer to
a somewhat larger value of
$\tilde{\omega}^2 (p_{Fn})$,   
without pion condensation. The
observed luminosities are labeled by dots. Possibilities of Fermi sea 
rearrangement and $\varrho^{\pm}$ condensation are ignored.
 \label{fig:hvh_nsf}}
\end{figure*}
The heat transport within the crust establishes homogeneous density profile
at times $\lsim (1\div10)$yr. 
After that time the subsequent cooling is determined
by simple relation $C_V \dot{T}=-L$, where $C_V =\sum_{i} C_{V,i}$ and
$L=\sum_{i} L_{i}$ are sums of the partial contributions
to the heat capacity (specific heat integrated over the volume)
and the luminosity (emissivity integrated over the volume).

The nucleon pairing gaps are rather purely known. Therefore one may vary them.
The {\em{"standard"}} and {\em{"nonstandard"}} scenarios of the  cooling
of NS of several selected masses for suppressed gaps
are demonstrated in the left panel of Fig. 5, \cite{SVSWW97}. 
Depending on the star
mass, the resulting photon luminosities are basically either too
high or too low compared to those given by  observations. 
Situation changes, if the MMU process (\ref{GU}) 
is included. Now, the cooling rates vary smoothly with
the star mass (see right panel of Fig. 5) such that the gap between
{\em{standard}} and  {\em{non-standard}} cooling scenarios
is washed out. More quantitatively,
by means of varying the NS mass between $(1 \div 1.6)~M_{\bigodot}$,
one achieves an agreement with a large number of observed data
points. This is true for a wide range of choices of the
$\widetilde{\omega}^{2}$ parameterization, independently whether 
pion (kaon) condensation can occur or not. Two
parameterizations presented in Fig. 5 with pion condensation for
$\varrho >3\varrho_0$ and without it
differ only in the range which is covered by the
cooling curves. The only point which does not agree with the
cooling curves belongs to the hottest pulsar PSR 1951+32. Other
three points which according to Fig. 5, right panel, are also not
fitted by the curves can be easily fitted by
slight changes of the model parameters. The high luminosity
of PSR 1951+32 may be due to internal heating processes, cf. \cite{SWWG96}.
\begin{figure*}
\centering\psfig{figure=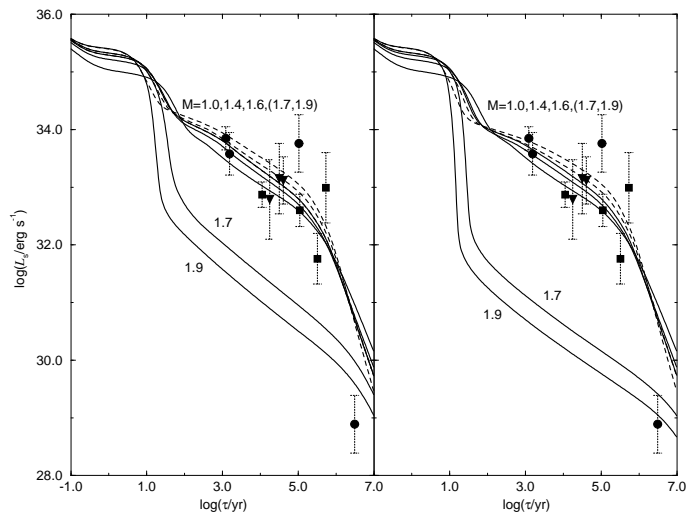,height=9cm,angle=-90}
\caption[]{Cooling of NS with different masses 
  constructed for the HV EOS \cite{SVSWW97}. 
The cooling processes are MU-VS86, PU (only solid curves), NPBF,
  PPBF, and DU (only in the right graph). The dashed curves refer to the
$M=1.7$ and $1.9M_{\odot}$ models without pion condensate. 
Possibilities of Fermi sea 
rearrangement and $\varrho^{\pm}$ condensation are ignored.
\label{fig:hvh_all_sf}}
\end{figure*}

We turn now to cooling simulations where the MU, NPBF, DU and PU
take place simultaneously.  Parameters of the pairing gaps are from Fig. 6 of
\cite{SVSWW97}. Fig. 6 shows the
cooling tracks of stars of different masses, computed for the
HV EQS. 
Very efficient at $T<T_c$ 
become to be NPBF processes 
which compete with MMU processes. The former prevail
for not too massive stars in agreement with estimation  
\cite{VS87}.
The DU process is taken into account in the
right graph, whereas it is neglected in the left graph. The solid
curves refer again to the $\tilde{\omega}^2$ parameterization with
phase transition to pion condensate, the dashed curves to the one
without phase transition (see Fig. 4). 
For masses in the range between 1.0 and 1.6
$M_\odot$, the cooling curves pass through most of the data points.
We again recognize a photon luminosity drop by more than two orders of
magnitude for the 1.7 $M_\odot$ mass star with pion condensate, due to
suppression of the pairing gaps in this case.
This drop is even larger if the DU is taken into account
(right graph). This allows to account for the photon luminosity of PSR
1929+10.

Thus, comparison with the observed luminosities shows that one
gets quite good agreement between theory and observations if one 
includes into consideration all available in-medium effects assuming
that the masses of the pulsars are different. 
We  point out that the description of these
effects is constructed in essentially the same manner for all the
hadron systems as NS,  atomic nuclei and heavy ion
collisions, cf. \cite{MSTV90,V93}.

\subsection{Neutrino opacity}

Important part of in-medium effects for description of neutrino 
transport  at initial
stage of NS cooling was discussed in
\cite{VS86,VSKH87,HKSV88,MSTV90}, where correlation effects, pion softening
and pion condensation, the latter for $\varrho >\varrho_{c\pi}$, were taken
into account. 
The neutrino/antineutrino  mean free paths
can be evaluated from the corresponding
kinetic equations via their 
widths $\Gamma_{\nu (\bar{\nu})} =-2\Im \Pi^R_{\nu (\bar{\nu})}$,
where $\Pi^R$ is the retarded self-energy,
or within 
the QPA for the nucleons they can be also estimated
via the squared matrix
elements of the corresponding reactions. Thereby the processes
which the most efficiently contribute to the emissivity 
are at early times (for $T\gsim 1$ MeV) responsible for the opacity.

In the above {\em{"nuclear medium cooling scenario"}} 
at $T>T_c$ the most essential contribution was from
MMU.
Taking into account of $NN$ correlations 
in the strong coupling vertices of two-nucleon processes like MMU and MNB
suppresses the rates, whereas   the softening of the
pion propagator essentially
enhances them. For rather massive NS
MOPE wins the competition.  
The mean free path of neutrino/antineurtino
in MMU processes is determined from the same diagrams (\ref{MMU-diag}) 
as the emissivity.
Its calculation (see (\ref{GU})) with the two first diagrams yields  
\begin{eqnarray}
\frac{\lambda_\nu^{MMU}}{R} \simeq  \frac{1.5 \cdot 10^{5}}{F_1 
(2\Gamma)^{8} T^{4}_9}
\left(\frac{\varrho_0
    }{\varrho}\right)^{10/3} \frac{m_N^4}{(m^{\ast}_n )^3 m^{\ast}_p }
\left[\frac{ \alpha \tilde{\omega}_{\pi^0}(p_{Fn})}{m_\pi}\right]^4
\left[\frac{ \alpha \tilde{\omega}_{\pi^\pm}(p_{Fn})}{m_\pi}\right]^4 .
\end{eqnarray}From the relation  $\lambda_\nu \simeq R$ follows  evaluation
of $T_{opac}$. With the only first diagram we get a simple estimate 
\begin{eqnarray}
T^{opac}_9 \simeq 11 \frac{\varrho_0}{\varrho}
\frac{\widetilde{\omega}^2 (p_{Fn}^2 )}{[4\gamma (g_{nn},0,p_{Fn})
\widetilde{\gamma}(g' ,0,p_{Fn}) ]^{1/2}}\frac{m_N}
{m_N^{\ast}}\,.
\end{eqnarray}
For averaged  value of the density $\varrho \simeq \varrho_0$ corresponding to
a medium-heavy NS
($<1.4 M_\odot$) with $\tilde{\omega}^2(p_{Fn})\simeq 0.8 m_{\pi}^2$, 
$\widetilde{\gamma}\simeq \gamma
\simeq (0.3\div0.4)$ 
we get $T_{opac}\simeq (1\div1.5)$ MeV
that is smaller then the value $T_{opac}\simeq 2$ MeV estimated with FOPE
\cite{FM79}.
For $\varrho \simeq 2\varrho_0$ that corresponds to a more massive
NS we evaluate  $T_{opac}\simeq (0.3\div0.5)$ MeV. 
Thus pion softening results in a substantial decrease of 
neutrino/antineutrino mean free
paths and the value of $T_{opac}$. 

The diffusion equation determines the characteristic time scale for the heat
transport of neutrinos from the hot zone to the star surface
$t_0 \sim R^2 C_V \sigma^{-1}T^{-3}/\lambda_\nu$ 
($\sigma$ is the Stefan--Boltzmann constant),
whereas it follows that $t_0\sim$10 min. for $T\simeq 10$ MeV and 
$\varrho \simeq \varrho_0$,
$\widetilde{\omega}^2 \simeq 0.8 m_\pi$, $\Gamma \simeq 0.4$, $m_N^\ast/m_N 
\simeq 0.9$, and $t_0 $ becomes as large as 
several hours for $\varrho \simeq (2\div3)\varrho_0$.
These estimates demonstrate 
that more massive NS cool down more slowly
at $T>T_{opac}$ and faster at subsequent times then the less massive stars.

Due to in-medium effects 
neutrino scattering cross section on the neutrons shown by the first diagram
(\ref{nu-scat}) requires the  $NN$ correlation factor 
\begin{eqnarray}\label{gam-nn}
\gamma^2_{n\nu}(\omega , q)=\frac{\gamma^2 (f_{nn} ,
\omega , q)+3g^2_{A}\gamma^2 (g_{nn}, \omega , q)}{1+3g^2_{A}},
\end{eqnarray}
as follows from (\ref{nn-cor}). 
This results in  a suppression of 
of the cross sections for $\omega < qv_{Fn}$ and in an enhancement for $\omega > qv_{Fn}$.
Neutrino scattering cross sections on the protons are modified by
\begin{eqnarray}\label{gam-pp}
\gamma^2_{p\nu}(\omega , q)=\frac{\kappa^2 (f_{np} , f_{nn},
\omega , q)+3g^2_{A}\gamma^2_{pp} (g_{nn}, \omega , q)}{1+3g^2_{A}},
\end{eqnarray}
that results in the same order of magnitude correction 
as given by (\ref{gam-nn}).

Also there is
a suppression of the $\nu N$ scattering 
and MNB reaction
rates for soft neutrinos ($\omega\lsim (3\div6)T$) 
due to multiple $NN$ collisions
\begin{eqnarray}\label{mult}
&\hspace*{5mm}...\unitlength8mm\begin{picture}(4.5,1)
   \put(0,-.35){\twocol}\end{picture}\hspace*{3cm}&
\end{eqnarray}
(LPM effect). 
One may estimate these effects  simply multiplying squared matrix elements
of the $\nu N$ scattering and the MNB processes  by the corresponding
suppression pre-factors
\cite{KV95}. Qualitatively one may use a general pre-factor
$C_0(\omega)=\omega^2 /[\omega^2+\Gamma_N^2]$,
where $\Gamma_N$ is the nucleon width and the $\omega$ is the energy
of $\nu$ or $\nu\bar{\nu}$ pair. In some works, see
\cite{RS95,HR97}, correction factor, like
$C_0$, was suggested at
an ansatz level. 
Actually one does not need any ansatze reductions. 
OTF, see \cite{KV95},  
allows to calculate the rates using the exact sum rule.
The modification
of the charged current processes due to LPM effect
is unimportant since the corresponding value of
$\omega$ is $\simeq p_{Fe}\gg \Gamma_N$. 

The main physical result we discussed is that in medium 
the reaction rates are essentially modified. A suppression arises
due to $NN$ correlations (for $\omega \ll kv_{FN}$) and 
infra-red pre-factors (coherence effects), and
an enhancement due to the pion softening (and $NN$ correlations for
$\omega \sim q$) and due to opening up of new efficient reaction channels. 
The  pion softening demonstrates that already
for densities $\varrho <\varrho_0$
the nucleon system begins to feel that it may have $\pi$
condensate phase transition for $\varrho
>\varrho_{c\pi}$, although this $\varrho_{c\pi}$ value might be 
essentially larger than $\varrho_{0}$ or even not achieved.

\section{The rate of radiation from dense medium. OTF}\label{How}

Perturbative diagramss are obviously irrelevant for calculation 
of in-medium processes and one should deal with dressed Green functions.
The QPA for fermions is applicable if the fermion
width is much less than all the typical energy scales essential in the problem 
($\Gamma_F \ll \omega_{ch}$). In calculation of the emissivities
of $\nu\bar{\nu}$ reactions
the minimal scale is $\omega_{ch}\simeq 6T$, averaged $\nu\bar{\nu}$
energy for MNB reactions. For MMU $\omega_{ch}\simeq p_{Fe}$. 
For radiation of soft
quanta of fixed energy $\omega <T$,  $\omega_{ch}\simeq\omega$.
Within the QPA for fermions, 
the  reaction rate with participation of the
fermion and  the boson is given by \cite{VS84,VS86}
\\
\\
\begin{equation}\label{o-n}
\parbox{30mm}{
\begin{fmfgraph*}(60,50)
\fmfleft{l}
\fmfrightn{r}{3}
\fmfpoly{full}{ul,ur,uo}\fmfpen{thick}
\fmfpoly{full}{dl,dr,do}
\fmfforce{(0.0w,0.8h)}{l}
\fmfforce{(1.3w,0.8h)}{r1}
\fmfforce{(0.7w,1.1h)}{r2}\fmfforce{(0.7w,1.3h)}{r3}
\fmfforce{(1.05w,0.0h)}{o}
\fmfforce{(0.4w,1.1h)}{o2}
\fmfforce{(0.4w,0.9h)}{uo}
\fmfforce{(0.5w,0.8h)}{ur}
\fmfforce{(0.3w,0.8h)}{ul}
\fmfforce{(0.95w,0.8h)}{dl}
\fmfforce{(1.15w,0.8h)}{dr}
\fmfforce{(1.05w,0.7h)}{do}
\fmf{fermion,width=1thick}{l,ul}
\fmf{fermion,width=1thick}{ur,dl}
\fmf{fermion,width=1thick}{dr,r1}
\fmf{boson,width=1thick,label=$\varphi$,label.side=right}{do,o}
\fmf{fermion,width=1thin}{r3,o2}
\fmf{fermion,width=1thin}{o2,r2}\fmf{dashes,width=1thin}{uo,o2}
\fmfforce{(1.0w,0.8h)}{ddr}
\fmfforce{(-0.1w,0.0h)}{ll1}\fmfforce{(-0.1w,1.0h)}{ll2}\fmf{plain}{ll1,ll2}
\fmfforce{(1.4w,0.0h)}{rl1}\fmfforce{(1.4w,1.0h)}{rl2}\fmf{plain}{rl1,rl2}
\fmflabel{$2$}{rl2}
\fmfforce{(-0.3w,0.5h)}{lll1}\fmfforce{(-0.2w,0.7h)}{lll3}
\fmfforce{(-0.2w,0.3h)}{lll2}\fmf{plain,width=2thin}{lll1,lll2}
\fmf{plain,width=2thin}{lll1,lll3}
\fmfforce{(1.6w,0.5h)}{rrr1}\fmfforce{(1.5w,0.7h)}{rrr3}
\fmfforce{(1.5w,0.3h)}{rrr2}\fmf{plain,width=2thin}{rrr1,rrr2}
\fmf{plain,width=2thin}{rrr1,rrr3}
\end{fmfgraph*}}\,\,\,\,\,\,\,\,\,\,.
\end{equation}
For equilibrium ($T\neq 0$) system there is the exact relation
\begin{equation}
(<\widehat{\varphi}^\dagger_2\widehat{\varphi}_1 
>)(p)=\I D^{-+}+\mid \varphi_c \mid^2 =\frac{A_B}{
\mbox{exp}(\frac{\omega}{T})-1}+\mid \varphi_c \mid^2 
,\,\,\,A_B=-2\mbox{Im}D^R ,
\end{equation}
where $(<\widehat{\varphi}^\dagger_2\widehat{\varphi}_1 
>)(p)$ means the Fourier component of the  corresponding non-equilibrium 
Green function and $\varphi_c$ is the mean field.
Thus the rate of the reaction is related to 
the boson spectral function $A_B$ and the
width ($\Gamma_B$) being determined by the corresponding
Dyson equation, see (\ref{pion-l}). $A_B$
is the delta-function at the spectrum branches
related to resonance processes, like zero sound. The poles associated with
the upper
branches do not contribute at small temperatures due to a tiny thermal
population of those branches. There is also a contribution to $\mbox{Im}D^R$
proportional to $\mbox{Im}\Pi^R$ given by the particle-hole diagram.
Within the QPA taking $\mbox{Im}$ part means the cut of the diagram.
Thus we show \cite{VS86} that this contribution is the same as that 
could be
calculated with the help of 
the squared matrix element of the two-nucleon process
\begin{equation}\label{sq}
\parbox{30mm}{
\begin{fmfgraph*}(60,50)
\fmfleftn{l}{2}
\fmfrightn{r}{4}
\fmfpoly{full}{ur,ul,uo}\fmfpen{thick}
\fmfpoly{full}{dr,do,dl}
\fmfforce{(0.0w,0.2h)}{l1}
\fmfforce{(0.0w,0.8h)}{l2}
\fmfforce{(1.2w,0.2h)}{r1}
\fmfforce{(1.2w,0.8h)}{r2}\fmfforce{(1.2w,1.05h)}{r3}
\fmfforce{(1.2w,1.25h)}{r4}
\fmfforce{(0.9w,1.05h)}{o2}\fmfforce{(0.9w,0.9h)}{o1}
\fmfforce{(0.4w,0.8h)}{ul}
\fmfforce{(0.6w,0.8h)}{ur}
\fmfforce{(0.5w,0.7h)}{uo}
\fmfforce{(0.4w,0.2h)}{dl}
\fmfforce{(0.6w,0.2h)}{dr}
\fmfforce{(0.5w,0.3h)}{do}
\fmf{fermion,width=1thick}{l2,ul}
\fmf{fermion,width=1thick}{ur,ddl}\fmf{fermion,width=1thick}{ddr,r2}
\fmf{fermion,width=1thick}{l1,dl}
\fmf{fermion,width=1thick}{dr,r1}
\fmf{boson,width=1thick,label=$D^R$,label.side=right}{do,uo}
\fmf{fermion,width=1thin}{r3,o2}
\fmf{fermion,width=1thin}{o2,r4}\fmf{dashes,width=1thin}{o1,o2}
\fmfpoly{full}{ddr,o1,ddl}\fmfforce{(0.8w,0.8h)}{ddl}
\fmfforce{(1.0w,0.8h)}{ddr}
\fmfforce{(-0.1w,0.0h)}{ll1}\fmfforce{(-0.1w,1.0h)}{ll2}\fmf{plain}{ll1,ll2}
\fmfforce{(1.3w,0.0h)}{rl1}\fmfforce{(1.3w,1.0h)}{rl2}\fmf{plain}{rl1,rl2}
\fmflabel{$2$}{rl2}
\end{fmfgraph*}}\,\,.
\end{equation}
This is precisely what one could expect  using optical theorem.
Thus unlimited series of all possible diagrams  with in-medium 
Green functions  
(see  (\ref{nu-scat}), (\ref{picond}), (\ref{res-pr}), (\ref{du})) 
together with two-fermion diagrams (as given by (\ref{MMU-diag})) 
and multiple-fermion
diagrams (like (\ref{mult})) would lead us to a double counting.
The reason is that permitting the boson width effects (and beyond the
QPA for fermions also permitting finite fermion widths)
the difference between one-fermion, two-fermion  and
multiple-fermion processes in medium is
smeared out. All the states are allowed and
participate in production and
absorption processes.
Staying with the QP picture for fermions, 
the easiest way to avoid mentioned double counting is to calculate 
the reaction rates according to (\ref{o-n}), i.e. 
with the help of the diagrams of the 
DU-like type,
which already include all the contributions of the two-nucleon origin.
Multiple $NN$ collision processes should be added separately. 
On the other hand, it is rather inconvenient to 
explicitly treat all one-nucleon
processes dealing with different specific quanta instead of  
using of the full $NN$ interaction amplitude whenever it is possible.
Besides, as we have mentioned,
consideration of open fermion legs is only possible within the
QPA for fermions since Feynman technique is not
applicable if Green functions of ingoing and outgoing 
fermions have widths (that in another language means possibility of
additional processes).
Thus the idea came \cite{VS87,KV95}
to integrate over all in-medium states allowing all
possible processes instead of specifying different special reaction channels.

In  \cite{VS87,KV95,KV99} it was shown that OTF in
terms of full non-equilibrium Green functions is an efficient tool to
calculate the reaction  rates including finite
particle widths and other in-medium effects. 
Applying this approach, e.g., to the antineutrino--lepton (electron, $\mu^-$
meson, or
neutrino)
production \cite{VS87}
we can express the transition probability in a direct reaction
in terms of the evolution operator $S$,
\begin{equation}\label{optfirst}
\frac{d{\cal W}^{\rm  tot}_{X\rightarrow \bar\nu l}}{d t}=
\frac{(1-n_l) d q_l^3 dq_{\bar{\nu}}^3}{(2\pi)^6\,4\,
\omega_l \,\omega_{\bar{\nu}}}
\,\sum_{\{X\}}\overline{<0|\, S^\dagger\, |\bar\nu l + X>\,<\bar\nu l +X|\,
S\, |0>}\,,
\end{equation}
where we presented explicitly the phase-space volume of 
$\bar\nu l$ states; lepton occupations of given spin, 
$n_l$, are put zero for
$\nu$ and $\bar{\nu}$
which are supposed to be radiated directly from the system (for $T<T_{opac}$). 
The bar denotes statistical averaging.
The summation goes over complete set of all possible intermediate
states $\{X\}$ constrained by the energy-momentum conservation. 
It was also supposed that electrons/muons can be treated
in the QPA, i.e. with zero widths. 
Then  there is no need (although possible) to consider
them  in intermediate reaction states. 
Making use of the smallness of the weak coupling, 
we expand the evolution operator as
$S\approx 1- \I \, \intop_{-\infty}^{+\infty} T\,\bigl\{V_W(x)\,
S_{\rm nucl} (x)\bigr\} d x_0
\,,$
where $V_W$ is the Hamiltonian of the weak interaction
in the interaction representation, $S_{\rm
nucl}$ is the part of the $S$ matrix corresponding to the nuclear
interaction, and $T\{...\}$ 
is the chronological  ordering
operator. After substitution 
into 
(\ref{optfirst}) and averaging over the arbitrary
non-equilibrium state of a nuclear system, there appear
chronologically ordered ($G^{--}$), anti-chronologically ordered ($G^{++}$) and
disordered ($G^{+-}$ and $G^{-+}$)
exact Green
functions. 

In  graphical form the general expression for the probability of
the lepton (electron, muon, neutrino) and anti-neutrino production is 
as follows
$$ \parbox{10mm}{\setlength{\unitlength}{1mm}
\begin{fmfgraph*}(30,10)
\fmfleftn{l}{2}
\fmfrightn{r}{2}\fmfpen{thick}
\fmfpoly{hatched,pull=1.4,smooth}{ord,oru,olu,old}
\fmfforce{(0.3w,.8h)}{olu}
\fmfforce{(0.7w,.8h)}{oru}
\fmfforce{(0.3w,0.2h)}{old}
\fmfforce{(0.7w,0.2h)}{ord}
\fmfforce{(0.25w,0.5h)}{ol}
\fmfforce{(0.75w,0.5h)}{or}
\fmf{fermion,width=1thin}{l1,ol}
\fmf{fermion,width=1thin}{ol,l2}
\fmf{fermion,width=1thin}{r2,or}
\fmf{fermion,width=1thin}{or,r1}
\fmflabel{$\bar\nu$}{l1}
\fmflabel{$l$}{l2}
\fmflabel{$\bar\nu$}{r2}
\fmflabel{$l$}{r1}
\fmfv{l=$+$,l.a=180,l.d=3thick}{ol}
\fmfv{l=$-$,l.a=0,l.d=3thick}{or}
\end{fmfgraph*}}
\,\,\, ,$$
\\
representing the sum of all closed diagrams ($-\I \Pi^{-+}$)
containing at
least one ($-+$) exact Green function. 
The latter quantity is especially important.
Various contributions from $\{X\}$ can be classified according
to the number $N$ of $G^{-+}$ lines in the diagram  
\\
\begin{eqnarray}\label{qp}
\frac{d{\cal W}^{\rm  tot}_{\bar\nu l}}{d t}=
\frac{(1-n_l)d^3 q_{\bar{\nu}} d q_l^{3} }{(2\pi )^6\,4\,\omega_{\bar{\nu}}\,\omega_l}
\left(\,\,
\parbox{27mm}{
\setlength{\unitlength}{1mm}
\begin{fmfgraph*}(27,10)
\fmfleftn{l}{2}
\fmfrightn{r}{2}\fmfpen{thick}
\fmfpoly{empty,pull=1.4,smooth,label=
$N=1$ }{ord,oru,olu,old}
\fmfforce{(0.3w,.8h)}{olu}
\fmfforce{(0.7w,.8h)}{oru}
\fmfforce{(0.3w,0.2h)}{old}
\fmfforce{(0.7w,0.2h)}{ord}
\fmfforce{(0.25w,0.5h)}{ol}
\fmfforce{(0.75w,0.5h)}{or}
\fmf{fermion,width=1thin}{l1,ol}
\fmf{fermion,width=1thin}{ol,l2}
\fmf{fermion,width=1thin}{r2,or}
\fmf{fermion,width=1thin}{or,r1}
\fmflabel{$\bar\nu$}{l1}
\fmflabel{$l$}{l2}
\fmflabel{$\bar\nu$}{r2}
\fmflabel{$l$}{r1}
\fmfv{l=$+$,l.a=180,l.d=3thick}{ol}
\fmfv{l=$-$,l.a=0,l.d=3thick}{or}
\end{fmfgraph*}
}
\,+\,
\parbox{27mm}{
\setlength{\unitlength}{1mm}
\begin{fmfgraph*}(27,10)
\fmfleftn{l}{2}
\fmfrightn{r}{2}\fmfpen{thick}
\fmfpoly{empty,pull=1.4,smooth,label=
$N= 2$ }{ord,oru,olu,old}
\fmfforce{(0.3w,.8h)}{olu}
\fmfforce{(0.7w,.8h)}{oru}
\fmfforce{(0.3w,0.2h)}{old}
\fmfforce{(0.7w,0.2h)}{ord}
\fmfforce{(0.25w,0.5h)}{ol}
\fmfforce{(0.75w,0.5h)}{or}
\fmf{fermion,width=1thin}{l1,ol}
\fmf{fermion,width=1thin}{ol,l2}
\fmf{fermion,width=1thin}{r2,or}
\fmf{fermion,width=1thin}{or,r1}
\fmflabel{$\bar\nu$}{l1}
\fmflabel{$l$}{l2}
\fmflabel{$\bar\nu$}{r2}
\fmflabel{$l$}{r1}
\fmfv{l=$+$,l.a=180,l.d=3thick}{ol}
\fmfv{l=$-$,l.a=0,l.d=3thick}{or}
\end{fmfgraph*}
}
\dots\right)\,\, .
\end{eqnarray}
\\
This procedure suggested in \cite{VS87}
is actually very helpful especially if the QPA
holds
for the fermions.  Then contributions of specific processes
contained in a closed diagram  can be made visible by cutting the
diagrams over the ($+-$), ($-+$) lines. In the framework of the 
QPA for the fermions $G^{-+}=2\pi \I n_{F}\delta 
(\varepsilon +\mu -\varepsilon^0_p -\mbox{Re}\Sigma^R (\varepsilon +\mu ,
\vec{p}))$ ($n_F$ are fermionic occupations, for equilibrium $n_F = 
1/[\mbox{exp} ((\varepsilon -\mu_F )/T)+1]$),
and the cut eliminating the energy integral thus requires
clear physical meaning.
This way
one establishes the correspondence between closed diagrams
and usual Feynman amplitudes although in general case of finite fermion width
the cut has only a symbolic meaning. 
Next advantage is that in the QPA 
any extra $G^{-+}$, since it is proportional to $n_F$,
brings a small $(T/\varepsilon_F )^2$
factor to the emissivity of the process. Dealing with small temperatures one
can restrict by the diagrams of the lowest order in $(G^{-+}G^{+-})$,
not forbidden by energy-momentum conservations, putting $T=0$ in all
$G^{++}$ and $G^{- -}$ Green functions. 
Each diagram in (\ref{qp}) represents a whole
class of perturbative diagrams of any order in the interaction
strength and in the number of loops.

Proceeding further we may explicitly 
decompose the first term in (\ref{qp}) as\\
\begin{equation}\label{opts}
\parbox{30mm}{
\setlength{\unitlength}{1mm}
\begin{fmfgraph*}(30,10)
\fmfleftn{l}{2}
\fmfrightn{r}{2}\fmfpen{thick}
\fmfpoly{empty,pull=1.4,smooth,label=
$N= 1$ }{ord,oru,olu,old}
\fmfforce{(0.3w,.8h)}{olu}
\fmfforce{(0.7w,.8h)}{oru}
\fmfforce{(0.3w,0.2h)}{old}
\fmfforce{(0.7w,0.2h)}{ord}
\fmfforce{(0.25w,0.5h)}{ol}
\fmfforce{(0.75w,0.5h)}{or}
\fmf{fermion,width=1thin}{l1,ol}
\fmf{fermion,width=1thin}{ol,l2}
\fmf{fermion,width=1thin}{r2,or}
\fmf{fermion,width=1thin}{or,r1}
\fmflabel{$\bar\nu$}{l1}
\fmflabel{$l$}{l2}
\fmflabel{$\bar\nu$}{r2}
\fmflabel{$l$}{r1}
\fmfv{l=$+$,l.a=180,l.d=3thick}{ol}
\fmfv{l=$-$,l.a=0,l.d=3thick}{or}
\end{fmfgraph*}
}=\,\,\,
\setlength{\unitlength}{1mm}
\parbox{30mm}{\begin{fmfgraph*}(30,10)
\fmfleftn{l}{2}
\fmfrightn{r}{2}\fmfpen{thick}
\fmf{fermion,width=1thin}{l1,ol,l2}
\fmf{fermion,width=1thin}{r2,or,r1}
\fmfforce{(0.2w,0.5h)}{ol}
\fmfforce{(0.8w,0.5h)}{or}
\fmfpoly{full}{old,olu,ol}
\fmfpoly{full}{or,oru,ord}
\fmfforce{(0.35w,0.8h)}{olu}
\fmfforce{(0.35w,0.2h)}{old}
\fmfforce{(0.65w,0.8h)}{oru}
\fmfforce{(0.65w,0.2h)}{ord}
\fmf{fermion,left=.4,tension=.5}{olu,oru}
\fmf{fermion,left=.4,tension=.5}{ord,old}
\fmflabel{$\bar\nu$}{l1}
\fmflabel{$l$}{l2}
\fmflabel{$\bar\nu$}{r2}
\fmflabel{$l$}{r1}
\fmfv{l=$+$,l.a=180,l.d=3thick}{ol}
\fmfv{l=$-$,l.a=0,l.d=3thick}{or}
\end{fmfgraph*}} \,\,\,+\,\,\,
\setlength{\unitlength}{1mm}
\parbox{30mm}{\begin{fmfgraph*}(30,10)
\fmfleftn{l}{2}
\fmfrightn{r}{2}\fmfpen{thick}
\fmf{fermion,width=1thin}{l1,ol,l2}
\fmf{fermion,width=1thin}{r2,or,r1}
\fmfforce{(0.2w,0.5h)}{ol}
\fmfforce{(0.8w,0.5h)}{or}
\fmfpoly{full}{old,olu,ol}
\fmfpoly{full}{or,oru,ord}
\fmfforce{(0.35w,0.8h)}{olu}
\fmfforce{(0.35w,0.2h)}{old}
\fmfforce{(0.65w,0.8h)}{oru}
\fmfforce{(0.65w,0.2h)}{ord}
\fmf{fermion,left=.2,tension=.5}{c,oru}
\fmf{fermion,right=.2,tension=.5}{c,olu}
\fmf{fermion,left=.2,tension=.5}{ord,d}
\fmf{fermion,right=.2,tension=.5}{old,d}
\fmfforce{(0.5w,0.0h)}{d}\fmfforce{(0.5w,1.0h)}{c}
\fmflabel{$\bar\nu$}{l1}
\fmflabel{$l$}{l2}
\fmflabel{$\bar\nu$}{r2}
\fmflabel{$l$}{r1}
\fmfv{l=$+$,l.a=180,l.d=3thick}{ol}
\fmfv{l=$-$,l.a=0,l.d=3thick}{or}
\end{fmfgraph*}} \,\,\, 
\,\,\, .\end{equation}
\\
The full vertex in the diagram  (\ref{opts}) of given sign is
irreducible with respect to  the
($+-$) and ($-+$) nucleon--nucleon hole lines. This means it contains
only the lines of given sign, all ($--$) or ($++$). Second diagram with
anomalous Green functions exists only for systems with pairing.
In the framework of the QPA 
namely these diagrams determine the proper DU 
and also NPBF processes calculated in \cite{VS87,SV87} within OTF. 
In the QP picture the contribution to DU process
vanishes 
for $\varrho <\varrho_{cU}$. 
Then the second term (\ref{qp}) comes into play
which within the same QP picture
contains two-nucleon
processes with one ($G^{-+}G^{+-}$) loop in intermediate states, etc.

The full set of diagrams for $\Pi^{-+}$  can be further explicitly decomposed
as series \cite{KV95} (from now in brief notations)
\begin{equation}\label{keydiagrams}
   \unitlength6mm\keydiagrams\; 
\end{equation}
Full dot is the weak coupling
vertex including all the
diagrams of one sign, $NN$ interaction block is the full block also of one sign
diagrams. The lines are full Green functions with the widths.
The most essential term is the
one-loop diagram (see (\ref{opts})),
which is positive definite, and including the fermion width
corresponds to the first term of the classical Langevin result, for details
see 
\cite{KV95}. Other
diagrams represent interference terms due to rescatterings.
In some simplified representations (e.g., as  we 
used within Fermi liquid theory) 
the 4-point functions (blocks of $NN$ interaction of given sign diagrams) 
behave like intermediate bosons (e.g. zero-sounds and 
dressed pions). In general
it is not necessary to consider different quanta dealing instead  
with the full $NN$ interaction (all diagrams of given sign).
For particle propagation in an external field, e.g. infinitely
heavy scattering centers (proper LPM effect), 
only the one-loop diagram remains, since 
one deals then with a genuine one-body problem.
In the quasiclassical limit for fermions (with small occupations $n_F$)
all the diagrams given by first line of
series (\ref{keydiagrams})
with arbitrary number of $"-+"$
lines are summed up leading to the diffusion
result, for details
see \cite{KV95}. For small momenta $q$ this
leads to a suppression factor of the form
$C={\omega^2}/{(\omega^2+\Gamma_x^2)}$, $\Gamma_x$ incorporates
rescattering processes.
In general case 
the total radiation rate is obtained by summation of  all
diagrams in (\ref{keydiagrams}). The value $-\I \Pi^{-+}$ determines 
the gain term in the generalized kinetic equation for $G^{-+}$, see 
\cite{IKV99,IKVV00}, 
that allows to use this method in non-equilibrium problems,
like for description of neutrino transport in semi-transparent region
of the neutrino-sphere of supernovae, as we may expect.

In the QP limit diagrams 1, 2, 4 and 5 of
(\ref{keydiagrams}) correspond to
the MMU and MNB processes related to a single in-medium scattering of two
fermionic QP and can be symbolically expressed as Feynman
amplitude (\ref{Feynman}a)\\
\begin{eqnarray} \label{Feynman}
   (a)&\hspace*{5mm}\unitlength8mm\begin{picture}(2.5,0.7)
   \put(0,-.35){\borndiag}\end{picture}\hspace*{4.7cm}&(b)
  \hspace*{5mm}\unitlength8mm\begin{picture}(2.5,0.7)
   \put(0,.2){\selfinsert}\end{picture}\nonumber\\[3mm]
  (c)&\hspace*{5mm}\unitlength8mm\begin{picture}(4.5,1)
   \put(0,-.35){\twocol}\end{picture}\hspace*{3cm}&(d)
  \hspace*{5mm}\unitlength6mm\begin{picture}(4.5,1)
   \put(0,-.55){\intrad}\end{picture}
\end{eqnarray}
The one-loop diagram in
(\ref{keydiagrams}) is particular, since its QP approximant 
in many cases vanishes as we have mentioned. 
However the full one-loop includes QP
graphs of the type (\ref{Feynman}b), which survive to the same order
in $\Gamma_N/\omega_{ch}$ as the other diagrams (therefore
in \cite{VS87} where QP picture was used this diagram was
considered as allowed diagram). In
QP series such a term is included in second diagram of
(\ref{qp}) although beyond the 
QPA it is included as the proper self-energy insertion to the one-loop 
result, i.e. in first term (\ref{qp}) \cite{KV95}.
In fact
it is positive definite and corresponds to the absolute square of the
amplitude (\ref{Feynman}a).  The other
diagrams 2, 4 and 5 of (\ref{keydiagrams}) describe the interference
of amplitude (\ref{Feynman}a) either with those amplitudes
where the weak coupling quantum
($l\bar{\nu}$ pair) couples to another leg 
or with one of the exchange diagrams. 
For neutral interactions diagram (\ref{keydiagrams}:2) is more
important than diagram 4  while this behavior reverses for charge
exchange interactions (the latter is important, e.g., for gluon
radiation from quarks in QCD transport due to color exchange
interactions).  Diagrams like 3 describe the interference terms due to
further rescatterings of the source fermion with others
as shown by (\ref{Feynman}c). 
Diagram (\ref{keydiagrams}:6) describes the production
from intermediate states and relates to the Feynman graph
(\ref{Feynman}d). For photons in the soft limit ($\omega \ll \varepsilon_F$)
this diagram (\ref{Feynman}d) gives a smaller contribution to the
photon production rate than the diagram (\ref{Feynman}a),  where
the normal bremsstrahlung contribution diverges like $1/\omega$
compared to the $1/\varepsilon_F$--value typical for the coupling to
intermediate fermion lines. For $\nu\bar{\nu}$ bremsstrahlung 
(\ref{Feynman}d) gives zero due to symmetry.
However in some cases
the process (\ref{Feynman}d) might be very important even in the soft
limit.  This is indeed the case for the MMU process considered above.
Some of the diagrams which are not
presented explicitly in (\ref{keydiagrams}) give more than two
pieces, if being cut, so they never reduce to the Feynman amplitudes. 
However in the QPA they give zero contribution \cite{KV95}.

With
$\Gamma_F \sim \pi^2T^2/\varepsilon_F$ for Fermi liquids, the criterion
$\Gamma_F \ll \omega_{ch} \sim T$ is satisfied 
for all
thermal excitations $\Delta \varepsilon \sim T\ll\varepsilon_F /\pi^2$. However
with the application to soft radiation this
concept  is no
longer justified. Indeed series of QP diagrams is not convergent
in the soft limit and there is no hope
to ever recover a reliable result by a finite number of QP diagrams
for the production of soft quanta. With {\em full Green functions},
however, one obtains a description that uniformly covers both the soft
($\omega\ll\Gamma_F $) and the hard ($\omega\gg\Gamma_F$) regimes.
In the vicinity of $\varrho_{c\pi}$ the quantity $\Gamma_F$
being roughly estimated in \cite{D75,D82}
as $\Gamma_F\propto \pi^2 \Gamma^2 Tm_\pi /\widetilde{\omega}$, and
coherence effects come into play.

In order to correct QP evaluations of different diagrams
by the fermion width effects for soft radiating quanta one can simply multiply
the QP results by different pre-factors \cite{KV95}.
E.g., comparing the one-loop result at non-zero $\Gamma_F$ with
the first non-zero diagram in the QPA
($\Gamma_F =0$ in the fermion
Green functions) we get 
\begin{equation}\label{corr0}
\unitlength8mm\begin{picture}(3.5,1)
   \put(0,0.15){\oneloopvertex}\end{picture}
   = C_0(\omega)
   \left\{\begin{picture}(3.5,1)\put(0,.15){\selfinsert}\end{picture}
   \right\}_{\mbox{QPA}},
\end{equation}
at small momentum $q$.
For 
the next order diagrams we have
\begin{eqnarray}
  \unitlength6mm\begin{picture}(3.5,1.)
  \put(0,0.15){\oneloopvertex}\put(1.625,0.15){\interaction}
  \end{picture}&=&C_1(\omega)\left\{\;
  \unitlength6mm\begin{picture}(3.5,1)
  \put(0,0.15){\oneloopvertex}\put(1.625,0.15){\interaction}
  \end{picture}\right\}_{\mbox{QPA}},\,
C_1(\omega)=\omega^2\frac{\omega^2-\Gamma^2_F }{(\omega^2+\Gamma^2_F )^2}\,,
\label{oneint}\label{corr1}\nonumber \\ 
  \cr \hspace*{1cm}&&\cr 
  \unitlength6mm 
  \begin{picture}(3.8,1.)
  \put(0,0.15){\doubleloop}\end{picture}&=&C_0(\omega)
  \left\{\unitlength6mm \begin{picture}(3.8,1.)
  \put(0,0.15){\doubleloop}\end{picture}\right\}_{\mbox{QPA}},\,\,\,
C_0(\omega)=\frac{\omega^2}{\omega^2+\Gamma_F^2}\; .
\label{corr2}
\end{eqnarray}
where factors $C_0$, $C_1$, ... 
cure the defect of the QPA for soft $\omega$. The factor $C_0$
complies with the replacement $\omega\rightarrow \omega+\I \Gamma_F $. A
similar factor is observed in the diffusion result, where
however the macroscopic relaxation rate $\Gamma_x$ enters, due to the
resummation of all rescattering processes.

Finally, we  demonstrated how to calculate the
rates of different reactions in dense equilibrium and
non-equilibrium matter and compared the results derived in closed 
diagram technique with those obtained in the standard technique of 
computing of the 
squared matrix elements. 

\section{Cooling of Hybrid Neutron Stars}\label{Quark}
Let us in addition review how the color superconducting quark matter, if exists
in interiors of massive NS, may affect the neutrino cooling of HNS, 
\cite{BGV00}. 
A detailed discussion of the neutrino emissivity of quark matter 
without taking into account of
the possibility of  the color superconductivity has  been given  
first in ref. \cite{I82}.  
In this work the quark direct Urca (QDU) 
reactions $d\rightarrow ue\bar{\nu}$ and $ue\rightarrow d {\nu}$ 
have been suggested as the most efficient processes.  
In the color superconducting matter
the corresponding expression for the  emissivity modifies as  
\begin{eqnarray}\label{neutr-DU1} 
\epsilon^{\rm QDU}_\nu \simeq 9.4 \times 
10^{26}\alpha_s (\varrho /\varrho_0 ) Y_e^{1/3}\zeta_{\rm QDU}~ 
T_{9}^6~ {\rm erg~cm^{-3}~s^{-1}}, 
\end{eqnarray} 
where due to the pairing the emissivity of QDU processes 
is suppressed by a factor, very roughly 
given by $\zeta_{\rm QDU} \sim \mbox{exp}(-\Delta_q /T)$. 
At $\varrho /\varrho_0\simeq 2$ the  
strong coupling constant is $\alpha_s \approx 1$ decreasing  
logarithmically at still higher densities, 
$Y_e =\varrho_e /\varrho$ is the electron fraction.  
If for somewhat larger density the electron fraction was too small  
($Y_e<Y_{ec}\simeq 10^{-8}$), 
then all the QDU processes  would be completely 
switched off \cite{DSW83} and the neutrino emission would be 
governed by two-quark reactions like the quark modified Urca (QMU) 
and the quark bremsstrahlung (QB) processes 
$dq\rightarrow uqe\bar{\nu}$ and $q_1 q_2 \rightarrow q_1 q_2 \nu\bar{\nu}$, 
respectively.  
The emissivities of QMU and QB processes are smaller than that for QDU
being suppressed by   
factor $\zeta_{\rm QMU} \sim \mbox{exp}(-2\Delta_q /T)$ for  
$T<T_{{\rm crit},q}\simeq 0.4~\Delta_q$. For  
$T>T_{{\rm crit},q}$ all the $\zeta$ factors are equal to unity.  
The modification of $T_{{\rm crit},q}(\Delta_q )$ relative to the standard  
BCS formula is due to the formation of correlations as, e.g., instanton- 
anti-instanton molecules.  
The contribution of the reaction $ee\rightarrow ee\nu\bar{\nu}$ to the
emissivity 
is very small \cite{KH99}, 
$\epsilon^{ee}_\nu \sim 10^{12}\, Y_e^{1/3} (\varrho /\varrho_0 )^{1/3} T_9^8~ 
{\rm erg~cm^{-3}~s^{-1}}$,   
but it can become important when quark processes are blocked out for large  
values of $\Delta_q/T$ in superconducting quark matter. 
 
For the quark specific heat \cite{BGV00} used 
expression of \cite{I82} being however
suppressed by the corresponding  $\zeta$ factor due to color superfluidity. 
Therefore gluon-photon and electron contributions play important role.
 
The heat conductivity of the matter is the sum of 
partial contributions 
$\kappa = \sum_{i}\kappa_{i},\,\,\, 
\kappa_i^{-1} = \sum_{j}\kappa_{ij}^{-1}~,$ 
where $i,j$ denote the components (particle species). 
For quark matter $\kappa$ is the sum of the partial conductivities of the 
electron, quark and gluon components 
$\kappa = \kappa_{e} + \kappa_{q}+\kappa_{g}$, 
where 
$\kappa_{e}\simeq \kappa_{ee}$ is determined by electron-electron scattering 
processes since in superconducting quark matter the partial contribution  
$1/\kappa_{eq}$ (as well as $1/\kappa_{gq}$ ) is additionally suppressed by a  
$\zeta_{\rm QDU}$ factor, as for the scattering on impurities in  
metallic superconductors.  Due to very small resulting value of $\kappa$
the typical time necessary for the heat to reach the star surface is large,
delaying the cooling of HNS.

EQS used in ref. \cite{BGV00} included a model
of the hadron matter,
region of mixed phase and of pure quark matter. A hard EQS
for the hadron matter was used, finite size effects
were disregarded in description of the mixed phase, and  
the bag constant $B$ was taken to be rather small that led to the presence of
a wide region of mixed and quark phases already 
for the HNS of the mass $M = 1.4~M_\odot$. On the other hand, absence of a
dense hadron region within this EQS
allowed to diminish uncertainties in description
of in-medium effects in hadronic matter suppressing them
according to that used in
the "standard scenario".

With the above inputs ref. \cite{BGV00} solved the evolution equation for the  
temperature profile.  
In order to demonstrate the influence of the size of the diquark and nucleon 
pairing gaps on the evolution of the temperature profile solutions were 
performed with different values of the quark and nucleon gaps. 
Comparison of the cooling evolution  
($\mbox{lg}~T_s$ vs. $\mbox{lg}~t$) of HNS of the mass
$M = 1.4~M_\odot$ is given in Fig. 7. 
The curves for $\Delta_q \gsim 1$ MeV are very close to  
each other  demonstrating typical large gap behaviour.  
The behaviour of the cooling curve for $t \leq 50 \div 100$ yr is in straight 
correspondence with the heat transport processes.  
The subsequent time evolution is governed by the processes in the hadronic  
shell and by a delayed transport within the quark core with a dramatically  
suppressed neutrino emissivity from the color superconducting region.  
In order to demonstrate this feature a calculation was performed  with 
the nucleon gaps ($\Delta_i (n), ~i= n,p$) being artificially  suppressed  
by a factor 0.2. 
Then up to $\mbox{lg}(t[{\rm yr}]) \lsim 4$ the behaviour of the cooling  
curve is analogous to the one would be obtained for pure hadronic matter. 
The curves labelled "MMU" show the cooling of hadron matter
with inclusion of appropriate medium modifications in the $NN$  
interaction.  
\begin{figure}
\centering\psfig{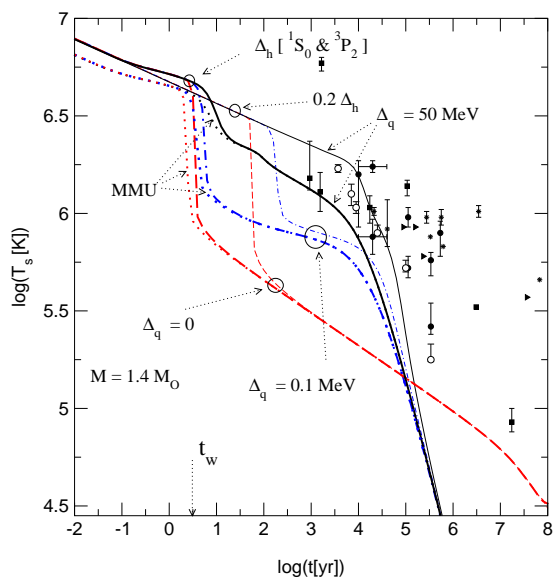}
\caption{Evolution of the surface temperature $T_s$ of HNS 
with $M=1.4 M_{\odot}$ for $T_s =(10 T_m )^{2/3}$, where $T$ is
in K, see \cite{T79}. Data  
are from \cite{SVSWW97} (full symbols) and from \cite{YLS99}  
(empty symbols), 
$t_w$ is the typical time which is necessary for the cooling 
wave to pass through the crust.}
\label{2scequal}
\end{figure} 
These effects have an influence on the cooling  
evolution only for $\mbox{lg}(t[{\rm yr}]) \lsim 2$ since the specific 
model EQS used does not allow for high nucleon densities in the 
hadron phase at given example of HNS of  $M = 1.4~M_\odot$.
The effect would be much more pronounced for larger star masses, a softer  
EQS for hadron matter and smaller values of the gaps in the  
hadronic phase. Besides, incorporation of finite size effects within
description of the mixed phase reducing its region should enlarge the 
size of the pure hadron phase.

The unique asymptotic behaviour at $\mbox{lg}(t[{\rm yr}]) \geq 5$ for all  
the curves corresponding to finite values of the quark and nucleon gaps is  
due to a competition between normal electron contribution to the specific 
heat and the photon emissivity from the surface since small exponentials  
switch off all the processes related to paired particles.  
This tail is very sensitive to the interpolation law $T_s =f(T_m)$ used to  
simplify the consideration of the crust.  
The curves coincide at large times due to the uniquely chosen relation  
$T_s \propto T_m^{2/3}$. 
 
The curves for $\Delta_q = 0.1$ MeV demonstrate an intermediate cooling 
behaviour between those for $\Delta_q=50$ MeV and $\Delta_q=0$. 
Heat transport becomes not efficient after first $5 \div 10$ yr.  
The subsequent $10^4$ yr evolution is 
governed by QDU processes and quark specific heat being only moderately 
suppressed by the gaps and by the rates of NPBF processes in the 
hadronic matter (up to $\mbox{lg}(t[{\rm yr}]) \leq 2.5$).  
At $\mbox{lg}(t[{\rm yr}]) \geq 4$ begins the photon cooling era. 
 
The curves for  normal quark matter ($\Delta_q =0$) are governed by 
the heat transport at times $t \lsim 5$ yr and then by QDU processes and the 
quark specific heat. The NPBF processes are important up to 
$\mbox{lg}(t[{\rm yr}])\leq 2$, the photon era is delayed up to  
$\mbox{lg}(t[{\rm yr}])\geq 7$.  
For times smaller than $t_w$ (see Fig. 7) the heat transport is delayed within 
the crust area \cite{LPPH91}. Since for simplicity this 
delay was disregarded in the
heat transport treatment, for such small times the curves should 
be interpreted as the $T_m(t)$ dependence scaled to guide the eye by the same 
law $\propto T_m^{2/3}$, as $T_s$.

For the CFL phase with large quark gap, which expected to exhibit  
the most prominent manifestations of color superconductivity in HNS and  
QCNS, \cite{BGV00} thus demonstrated an essential delay 
of the cooling during the first  
$50 \div 300$ yr (the latter for QCNS)
due to a dramatic suppression of the heat conductivity in  
the quark matter region.  
This delay makes the cooling of HNS and QCNS not as rapid as 
one could expect when ignoring the heat transport. 
In HNS compared to QCNS (large gaps) there is
an additional delay of the subsequent  
cooling evolution which comes from the processes in pure hadronic matter. 

In spite of that we find still too fast cooling for those objects compared to
ordinary NS. 
Therefore, with the CFL phase of large quark gap it seems   
rather difficult to explain the majority of the presently known 
data both in the cases of the HNS and QCNS, whereas in the case 
of pure hadronic stars the available data are much better fitted even
within the  
same simplified model for the hadronic matter. 
For 2SC (3SC) phases one may expect analogous behaviour to that demonstrated 
by $\Delta_q =0$ since QDU processes on unpaired quarks are then allowed,
resulting in a fast cooling.    
It is however not excluded that new observations may lead to 
lower surface temperatures for some supernova remnants and will be 
better consistent with the model which also needs further improvements. 
On the other hand, if future observations will show very large temperatures  
for young compact stars they could be interpreted as a manifestation of  
large gap color superconductivity in the interiors of these objects.

{\bf Concluding}, the {\em{ "nuclear medium cooling scenario"}}
allows easily to achieve agreement with existing data. However there remains
essential uncertainty in quantitative predictions due to a poor
knowledge, especially, 
of the residual interaction treated  above in an economical 
way within a
phenomenological Fermi liquid model which needs further essential
improvements. As for color superconductivity in HNS, QCNS, and QS,
characterizing by large diquark pairing gaps, we did not find an appropriate
fit of existing $X$ ray data. 
Situation will be changed if more cold, old objects will be observed.
Also, if young hot objects ( on scale $\sim 10^2~$ yr.) 
will be observed it could be interpreted as a signature of 
CFL phase in these objects.

{\bf{ Acknowledgement}}:
Author appreciates the hospitality and support of GSI Darmstadt
and ECT* Trento. He thanks E. Kolomeitsev for discussions.

\end{fmffile}

\end{document}